\documentclass[a4paper,11pt]{article}
\usepackage{jheppub}
\usepackage{cancel}
\usepackage{booktabs}
\usepackage{graphicx}

\def\eqref#1{{Eq.\!~(\ref{#1})}}
\def\figref#1{{Fig.\!~\ref{#1}}}
\def\secref#1{{Sec.\!~\ref{#1}}}

\title{\boldmath Finite Formation Time Antenna Radiation: Color Spectators Break Single-Emitter BDMPS-Z }

\author[a,b]{Fabio~Dom\'inguez}
\author[a]{, Pablo Guerrero-Rodr\'iguez}
\author[a,c]{, Carlos A.\ Salgado}

\affiliation[a]{Instituto Galego de F\'{\i}sica de Altas Enerx\'{\i}as IGFAE,
Universidade de Santiago de Compostela,\\ 15782 Santiago de Compostela, Galicia-Spain}
\affiliation[b]{CPHT, CNRS, Ecole polytechnique, Institut Polytechnique de Paris, 91120 Palaiseau, France}
\affiliation[c]{Axencia Galega de Innovaci\'on (GAIN), Xunta de Galicia, Galicia, Spain}

\emailAdd{fabio-alejandro.dominguez-gonzalez@polytechnique.edu}
\emailAdd{pablo.guerrero@usc.es}
\emailAdd{carlos.salgado@usc.es} 

\abstract{We compute the direct contribution to the squared matrix element for the in-medium soft gluon emission off a color-singlet quark-antiquark pair. To do this, we incorporate medium interactions that take place during the finite formation time of the antenna, thereby accounting for a largely neglected source of modifications with respect to the vacuum baseline. As a consequence, non-trivial correlations with the spectator (i.e.\ non-emitting) leg of the antenna emerge, greatly increasing the complexity of the result even at leading-$N_c$ order. The observed modifications entail a significant departure from the limit of instantaneous antenna formation, where the calculation effectively reduces to the well-known BDMPS-Z spectrum.}

\usepackage[dvipsnames]{xcolor}

\usepackage{soul}

\definecolor{greenish}{RGB}{30., 144., 4.}
\definecolor{blueish}{RGB}{11., 96., 172.}
\definecolor{redish}{RGB}{230., 0., 14.}
\definecolor{lcolor}{rgb}{0.,0.0,0.}
\definecolor{citcolor}{rgb}{0,0.,0.5}
\definecolor{mygray}{gray}{0.6}

\newcommand{\bm}[1]{\boldsymbol{#1}}

\newcommand{\diff}{{\rm d}}

\begin{document}
\maketitle
\flushbottom

\vspace{-1.54em}
\section{Introduction}
\label{sec:intro}
Under extreme conditions of temperature and density, Quantum Chromodynamics (QCD) predicts the formation of a novel state of matter in which quarks and gluons are not confined within hadrons. This phase, known as the Quark-Gluon Plasma (QGP), filled the universe in the earliest moments after the Big Bang, and it can be created in the laboratory through ultra-relativistic collisions of heavy nuclei. These experiments are currently conducted at the Relativistic Heavy Ion Collider (RHIC) and the Large Hadron Collider (LHC) with the aim to determine and quantify the properties of the QGP. However, given its extremely short lifetime (of the order of 10 yoctoseconds, $\sim\!10^{-23}\,\text{s}$), this substance cannot be manipulated or probed externally. Instead, one must rely on highly energetic probes generated in the same event, generically labeled as hard probes.

Hard probes originate from rare high-momentum exchanges between partons in the incoming nuclei, and propagate through the surrounding medium without thermalizing with it. One of the main examples are jets: collimated sprays of hadrons that emerge from cascades of particles emitted by a single hard parton. When these particles traverse a QCD medium, they interact with it, thus eventually fragmenting into jets that exhibit different features than those that develop in the vacuum. This effect is referred to as jet quenching, and it manifests through a wide range of experimental signatures (see e.g.\ \cite{Cao:2020wlm} for a review).
Some of the most prominent, such as the suppression of the high-$p_T$ spectrum \cite{PHENIX:2004vcz,STAR:2005gfr,PHOBOS:2004zne,BRAHMS:2004adc,PHENIX:2007yjc,CMS:2012aa,ATLAS:2015qmb,ALICE:2010yje}, the modification of jet yields \cite{CMS:2021vui,ATLAS:2014ipv,ALICE:2015mjv,CMS:2016uxf,STAR:2020xiv,ATLAS:2018gwx,ALICE:2019qyj,ALICE:2015mdb,ALICE:2018vuu,STAR:2023pal,ALICE:2023waz,ATLAS:2012tjt} and the enhancement of di-jet asymmetry \cite{ATLAS:2010isq,CMS:2011iwn,STAR:2016dfv}, can be attributed to a depletion of the jet energy into softer particles emitted outside the jet cone (energy loss). A precise theoretical description of this mechanism allows us to extract properties of the QGP through comparisons with experimental data. The earliest formulations of such a description were provided by Bjorken \cite{Bjorken1982EnergyLO}, Gyulassy and Wang \cite{Gyulassy:1993hr}; and later developed by Baier, Dokshitzer, Mueller, Peigné, Schiff \cite{Baier:1996sk,Baier:1996kr}, and Zakharov \cite{Zakharov:1996fv,Zakharov:1997uu} into what is known as the BDMPS-Z formalism \cite{Baier:1998kq}.

The focus of the BDMPS-Z formalism is the energy loss experienced by a highly energetic parton traversing a very dense medium. In this limit, the parton suffers multiple elastic scatterings with the medium constituents, which are treated as an ensemble of static, independent scattering centers. These interactions are resummed to all orders in the close-to-eikonal approximation via a path integral\footnote{Alternative formulations of this resummation process include the opacity expansion scheme (introduced by Gyulassy, Levai, Vitev \cite{Gyulassy:1999zd,Gyulassy:2000fs,Gyulassy:2000er} and independently by Wiedemann \cite{Wiedemann:2000za}) and the Hard Thermal Loop field theory (developed by Arnold, Moore and Yaffe \cite{Arnold:2001ba,Arnold:2001ms,Arnold:2002ja}).}, which in most applications is computed in the multiple soft scattering regime, e.g.\ \cite{Wiedemann:2000za,Wiedemann:2000tf,Salgado:2003gb}. The gluon spectrum obtained in this framework is finite in the collinear limit.
This is in stark contrast with the probability of gluon emission in the vacuum, which, owing to the collinear singularity, is logarithmically enhanced at small angles. Medium-induced emissions are thus expected to happen at relatively large angles, which leads to a degradation of the jet energy. However, a complete description of in-medium jet evolution demands that we take into account coherence effects between multiple emitters, as these could in principle suppress large-angle emissions. This is well known from vacuum QCD cascades, where destructive interference between different emitters leads to a progressive reduction of the angle of each successive emission (angular ordering) \cite{MUELLER1981161,PismaZhETF.33.285,Dokshitzer:1991wu}. It is therefore essential to understand precisely how the presence of a medium modifies this picture.

An ideal laboratory to investigate color coherence effects is the quark-antiquark dipole emitted by a highly virtual photon, also known as color-singlet QCD antenna. The in-medium soft gluon emission probability off such a system (as well as the color-octet antenna, which is emitted by a virtual gluon) has been extensively studied under different approximations. On a first approach, the medium was assumed to be relatively dilute, limiting the interaction with the antenna to a single scattering \cite{Mehtar-Tani:2010ebp}. In addition, the soft limit was applied, which implies that the gluon is emitted outside of the medium. This calculation would be revisited later in \cite{Mehtar-Tani:2011lic} for finite gluon energies. In \cite{Mehtar-Tani:2011hma} the antenna spectrum was obtained in the case of a dense medium, by resumming multiple scatterings within the soft limit and the strict eikonal approximation. The problem was then approached by relaxing the soft limit and including close-to-eikonal contributions within the aforementioned path integral formalism \cite{Mehtar-Tani:2011vlz,Casalderrey-Solana:2011ule,Mehtar-Tani:2012mfa}. Cumulatively, these studies lead to the conclusion that, for sufficiently small emission angles and short time scales, the background medium is unable to resolve the individual color charges carried by each leg of the antenna.
In this scenario, the interaction would happen coherently with the whole system, which, being a color singlet, is effectively invisible to the medium. This results in a vacuum-like (i.e.\ angle-ordered) emission pattern. Conversely, for a sufficiently wide antenna (or after an antenna has propagated for long enough inside the medium), each of its prongs becomes able to exchange color with the background independently, thus becoming decorrelated and spoiling the interference effects that dictate the radiation pattern in the vacuum. This phenomenon is known as decoherence, as it leads (in the extreme case) to the antenna spectrum becoming the incoherent sum of the spectra of quark and antiquark. As this happens, the radiation phase space effectively opens up to allow large-angle emissions \cite{Mehtar-Tani:2011vlz,Casalderrey-Solana:2011ule,Mehtar-Tani:2012mfa}.

The aforementioned studies were performed under the assumption that the splitting of the virtual photon into a quark-antiquark dipole takes place instantaneously, thus neglecting the medium interactions that might happen during the formation of the antenna itself. However, it is unclear whether the intuitive picture of color decoherence discussed above
remains valid when one removes this approximation. This question was first explored in \cite{Dominguez:2019ges,Isaksen:2023nlr}. In these works, the splitting of the photon into a quark-antiquark pair is assumed to take place within a finite formation time, during which interactions with the background medium are allowed. Their results showed the existence of phase space regions populated by vacuum-like emissions, with medium effects emerging for wide-angle splittings. Then, in \cite{Abreu:2024wka} a similar setup was applied to the calculation of the one-gluon emission spectrum, where, by virtue of the soft limit, the gluon is assumed to be emitted outside of the medium.
In the present work, we take a further step by considering the in-medium one-gluon emission by a singlet quark-antiquark antenna generated within a finite formation time. Although we still operate within the soft limit, we relax it enough to assume that the gluon emission takes place inside the medium. By doing this, we allow for the non-emitting leg of the pair (which is largely irrelevant in the instantaneous antenna formation limit) to participate in the color exchange with the medium, thereby giving rise to nontrivial correlations that modify the spectrum of the emitting prong. In order to showcase and study this effect, we focus our discussion on the direct contribution to the squared matrix element of the process, leaving a complete analysis of the gluon spectrum (including interference terms) for future work. 

This paper is organized as follows. In \secref{sec:setup} we briefly outline the BDMPS-Z formalism and illustrate it by computing the squared matrix element for a gluon emission off a single hard quark within a QCD medium. This allows us to introduce the specific approximations applied in this work, as well as our notation. In \secref{sec:calculation} we apply this formalism to compute the direct contribution to the squared matrix element of the in-medium gluon emission off a singlet quark-antiquark antenna. We also present the leading terms of its large-$N_c$ expansion and discuss their physical interpretation.
Then, in \secref{sec:limits} we examine two limiting cases of the obtained results, and use them to discuss the role of the spectator leg of the antenna in the process. Lastly, we present our conclusions and future prospects in \secref{sec:conclusions}. In order to lighten the notation and constrain the length of the main body of the manuscript, much of the calculational details (including the involved color algebra transformations required to obtain our main results) are left for the appendices. Specifically, in appendix~\ref{app:medav} we outline our treatment of the color structure of the medium averages in the fundamental representation. Then, in appendix~\ref{app:adjoint} we show how some propagators appearing in the results of the previous appendix are transformed back to the adjoint representation, thus arriving at our final expressions. In appendix~\ref{app:limit} we discuss the interpretation and relevance of a subset of terms within the leading-$N_c$ order of our result.

\section{The BDMPS-Z framework}
\label{sec:setup}
In order to showcase our notation and introduce the physical setup of the calculation, we consider the case of a highly energetic massless quark emitting a soft gluon within a dense QCD background medium. As they propagate, both particles experience multiple soft scatterings with the medium constituents, which are modeled as a collection of static color charges. The BDMPS-Z framework formulates this process as a path integral that resums these interactions to all orders along their trajectories. In this approach, the propagation of a parton from transverse position $\bf{x}$ at light-cone time $t$ to transverse position $\bf{y}$ at light-cone time $\bar{t}$ is described by the following Green's function:
\begin{align}\label{dressedprop}
{\cal G}^{ij}(\bar{t},{\bf y};t,{\bf x}|E)=\int^{{\bf r}(\bar{t})={\bf y}}_{{\bf r}(t)={\bf x}} {\cal D} {\bf r}(s) \, \exp\left\{ \frac{iE}{2}\int \diff s \left( \frac{\diff{\bf r}}{\diff s}\right)^2\right\} V^{ij}(\bar{t},t;\left[{\bf r}(s)\right]),
\end{align}
where the light-cone energy $E$ is assumed to remain fixed throughout. Essentially, this object captures the interaction with the medium as a convolution of the Feynman propagator of a free particle (which encodes the random transverse diffusion experienced by the parton in the close-to-eikonal approximation) with a Wilson line (which accounts for its precession in color space from $i$ to $j$ along a certain transverse path ${\bf r}$):
\begin{align}
V^{ij}(\bar{t},t;\left[{\bf r}(s)\right])={\cal P} \exp \left\{ ig \int^{\bar{t}}_{t}\diff s\,{\cal A}^{-a}(s,{\bf r}(s))T^a\right\}^{ij}\!\!\!\!.
\end{align}
Here, $T^a_{ij}$ is the color matrix in the corresponding representation and ${\cal A}^{-a}$ is the background medium field. This field is generated by a stochastic ensemble of color charges whose configuration we must average over when computing observables. In order to perform this procedure analytically, we assume that the medium fields obey Gaussian statistics:
\begin{align}\label{2pf}
\left\langle {\cal A}^{-a}(t,{\bf x} ) {\cal A}^{-b}(\bar{t},{\bf y})\right\rangle=n(t)\delta^{ab}\delta(t-\bar{t})\gamma(\bf{x}-\bf{y}),
\end{align}
where $n(t)$ is the number density of scattering centers and $\gamma({\bf x})$ is related to the elastic scattering rate.
This expression implies that all correlations are local in color and time.
In this work we adopt the same limit employed in e.g.\ \cite{Mehtar-Tani:2017ypq}, where quarks and antiquarks are assumed to be so energetic that their trajectories can be approximated with completely straight lines. This amounts to taking $E\!\rightarrow\!\infty$ in \eqref{dressedprop}, which recovers the strict eikonal propagator along the classical path between ${\bf x}$ and ${\bf y}$, i.e.\ the (tilted) Wilson line:
\begin{align}
{\cal G}^{ij}(\bar{t},{\bf y};t,{\bf x}|\infty)\rightarrow V^{ij}(\bar{t},t;\left[{\bf r}_{\text{cl}}(s)=\!{\bf p}(s-t)/E\right]),
\end{align}
where ${\bf p}$ is the transverse momentum. The `dressed' propagator from \eqref{dressedprop} will be applied only to describe the radiated gluon, which is assumed to be much softer than the parent parton and therefore sensitive to the sub-eikonal effects encoded in the path integral. These propagators thus carry adjoint Wilson lines, which we denote $U^{ab}$. Henceforth, the notation $V^{ij}$ will be reserved exclusively for Wilson lines in the fundamental representation.

\begin{figure}
\centering
\includegraphics[width=0.49\textwidth]{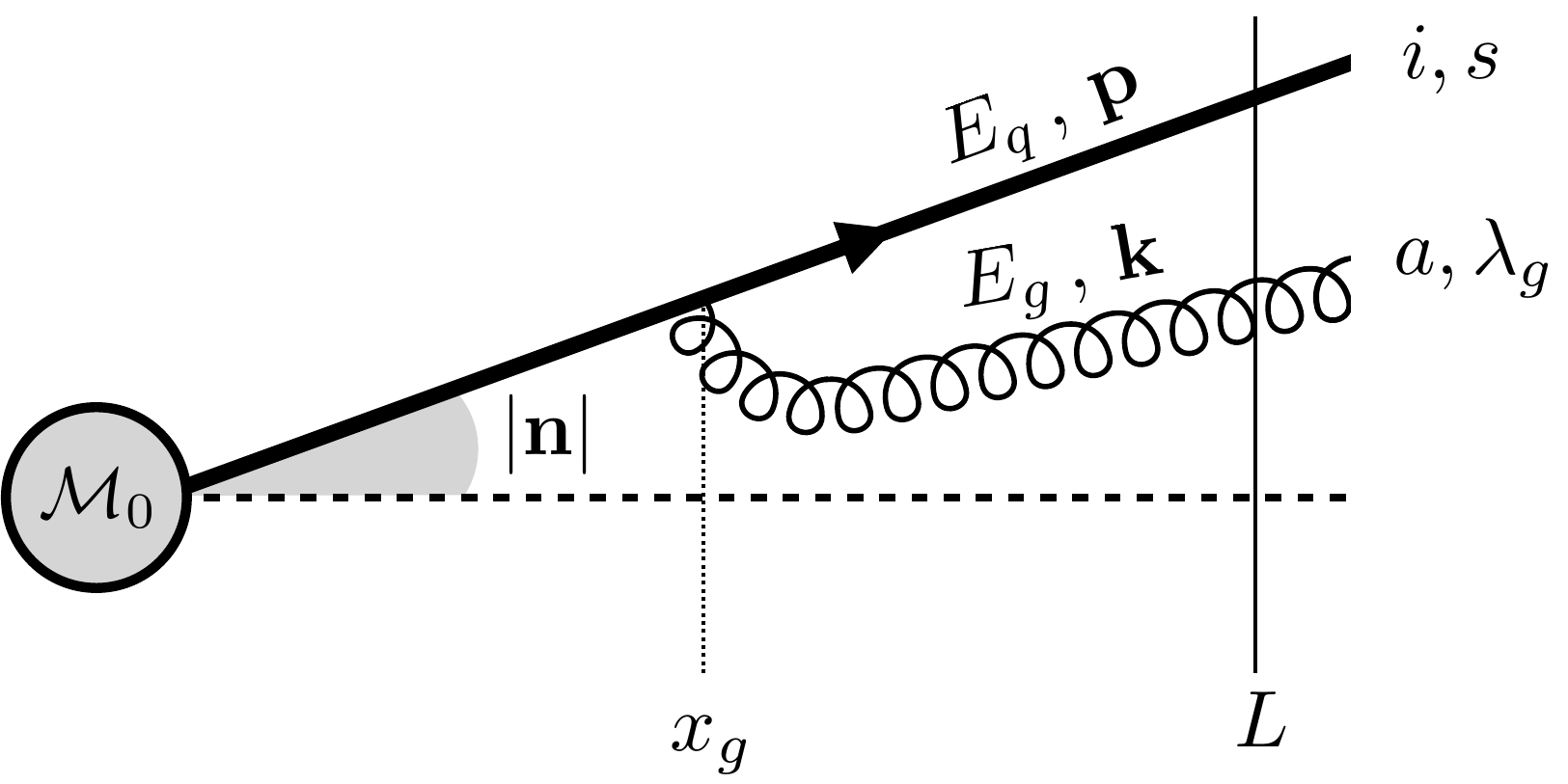}
\caption{Diagrammatic representation of the amplitude of a highly energetic quark emitting a gluon within a dense QCD medium of length $L$. We emphasize our choice of a quark trajectory that is non-constant in the transverse plane, following instead a straight line tilted at an angle approximately equal to $|{\bf n}|=|{\bf p}|/E$.}
\label{fig:bdmps}
\end{figure}

Having established our formalism, we put it into practice by computing the squared matrix element of the process represented in \figref{fig:bdmps}: a quark (with final energy $E_q$ and transverse momentum ${\bf p}$) emitting a soft gluon (with final energy $E_g$ and transverse momentum ${\bf k}$) at light-cone time $x_g$. For our purposes it will suffice to consider only the `in' contribution, where the gluon emission happens inside a medium of length $L$. By making use of the Feynman rules for in-medium processes (see e.g.\ \cite{Barata:2024bqp,Blaizot:2012fh}) and following the derivation steps from \cite{Mehtar-Tani:2011vlz,Casalderrey-Solana:2011ule}, one obtains the corresponding amplitude as\footnote{Throughout this manuscript we will denote transverse integrations as $\int_{\bf k}=\int \diff^2{\bf k}/(2\pi)^2$ for momenta and $\int_{\bf x}=\int \diff^2{\bf x}$ for positions.}:
\begin{align}\label{amp1}
\hspace{-0.25em}i{\cal M}^{i,s;A,\lambda_g}({\bf k}, {\bf p})\!=&\,\frac{ig}{2(E_q+E_g)} e^{i\frac{{\bf k}^2}{2E_g}L}\!\!\int^L_{0}\!\diff x_g\, e^{i\frac{E_g}{2}{\bf n}^2x_g}\!\! \int_{{\bf k}_1} e^{-i\,{\bf n}\cdot{\bf k}_1x_g}e^{-i\frac{{\bf p}^2}{2(E_q+E_g)}x_g}\Gamma^{ss_1}_{\lambda_g}\left[p;k_1\right]\nonumber\\
&{\cal G}^{ab}(L,{\bf k};x_g,{\bf k}_1|E_g)\left[V(L,x_g;[{\bf r}(s)])t^bV(x_g,0;[{\bf r}(s)])\right]^{il}\!{\cal M}^{l,s_1}_0(p),
\end{align}
where $t^b$ is the color generator in the fundamental representation and $s$, $\lambda_g$ are the spin and polarization of quark and gluon, respectively.
In the above expression, ${\cal M}^{l,s_1}_0\!(k_1+p)\approx{\cal M}^{l,s_1}_0\!(p)$ represents the local current encoding the creation of a quark with color $l$, spin $s_1$, energy $(E_q+E_g)$ and transverse momentum $({\bf p}+{\bf k}_1)\approx {\bf p}$. Note that, despite the quark's kinematics being modified in the radiation process, we treat it as completely recoil-less, thus having it traverse the same path before and after emission: ${\bf r}(s)\!=\!{\bf p}s/E_q\equiv{\bf n}(s-t)$.
The quark-gluon vertex, which we assume to be unaltered by the medium, reads:
\begin{align}
\Gamma^{ss_1}_{\lambda_g}\left[p;k_1\right]\,&=-\frac{2}{z_g\sqrt{1-z_g}}\delta_{ss_1}\left(\delta_{\lambda_g s_1}+(1-z_g)\delta_{\lambda_g-s_1}\right)((1-z_g){\bf k}_1-z_g{\bf p})\cdot\bm{\epsilon}^{*}_{\lambda_g}\nonumber\\
&\equiv \gamma_{\lambda_g}^{ss_1}(z_g)((1-z_g){\bf k}_1-z_g{\bf p})\cdot\bm{\epsilon}^{*}_{\lambda_g}\approx\gamma_{\lambda_g}^{ss_1}(z_g)\left({\bf k}_1-E_g{\bf n}\right)\cdot\bm{\epsilon}^{*}_{\lambda_g}\label{qgvertex},
\end{align}
where $z_g$ is the energy-sharing fraction, $z_g\!=\!E_g/(E_q\!+\!E_g)$.
By substituting this expression into \eqref{amp1} and inverse Fourier-transforming the gluon propagator to coordinate space, one eventually arrives at:
\begin{align}\label{eq:third1}
i{\cal M}({\bf k}, {\bf p})=&\,\frac{ig}{2(E_q+E_g)}\int^L_{0}\diff x_g \, e^{i\frac{E_g}{2}{\bf n}^2x_g}\int_{{\bf z}} e^{-i{\bf k }\cdot {\bf z}} \gamma^{\lambda_g}_{ss_1}(z_g)(i\partial_{{\bm \alpha}}\!-\!E_g{\bf n})\!\cdot\bm{\epsilon}^{*}_{\lambda_g}\nonumber\\
&{\cal G}^{ab}(L,{\bf z};x_g,{\bm \alpha})|_{\bm{\alpha}={\bf n}x_g}\left[V(L,x_g)t^bV(x_g,0)\right]^{il}\!{\cal M}_0^{l,s_1}.
\end{align}
Note that we will simplify our notation whenever no confusion can arise. In the above expression, the integration over ${\bf k}_1$ was absorbed into the (inverse) Fourier-transformed gluon vertex, $(i\partial_{{\bm \alpha}}\!-\!E_g{\bf n})$. By multiplying \eqref{eq:third1} with its complex conjugate and performing an average (sum) over initial (final) quantum numbers, one obtains:
\begin{align}\label{aux1}
\langle|{\cal M}|^2\rangle=&\frac{g^2}{E_g^2}2\text{Re}\int^L_{0}\diff x_g\int^L_{x_g}\diff\bar{x}_g \, e^{i\frac{E_g}{2}{\bf n}^2(x_g-\bar{x}_g)}\int_{{\bf z}\,\bar{\bf z}} e^{-i{\bf k }\cdot({\bf z}- \bar{\bf z})}(i\partial_{{\bm \alpha}}\!-\!E_g{\bf n})\!\cdot\!(-i\partial_{{\bm \beta}}\!-\!E_g{\bf n})\nonumber\\
&\frac{1}{2N_c}\left\langle {\cal G}^{\dagger b'a}(\bar{x}_g,{\bm \beta};L,\bar{\bf z})\,{\cal G}^{ab}(L,{\bf z};x_g,{\bm \alpha})\,U^{\dagger bb'}(x_g,\bar{x}_g)\right\rangle|_{\substack{\bm{\alpha}={\bf n}x_g\\ \bm{\beta}={\bf n}\bar{x}_g}}\,,
\end{align}
where we omitted the factor $|{\cal M}_0|^2$ for simplicity. Note that we combined the contributions where $x_g\!<\!\bar{x}_g$ and $\bar{x}_g\!<\!x_g$, which gives rise to the $2\text{Re}$ operator. Further, to arrive at this expression we applied the identity $\text{Tr}\{V^{\dagger}(x_g,\bar{x}_g)t^{b'}V(\bar{x}_g,x_g)t^b\}=\frac{1}{2}U^{b'b}(\bar{x}_g,x_g)$.

We now move on to the average over medium configurations. Since this operation is assumed to be local both in time and color, correlators like the one featured in \eqref{aux1} can be strategically factorized into distinct time regions, which must be subsequently projected onto color-singlet states, i.e.\ traces of Wilson lines. To do this, one needs to split the particle propagators into regions. Doing this in the case of the `dressed' gluon propagators requires that we integrate over the intermediate transverse position, which we call ${\bf z}_1$:
\begin{align}\label{Gsplit}
{\cal G}^{ab}(L,{\bf y};x_g,{\bf x})=\int_{{\bf z}_1}{\cal G}^{ac}(L,{\bf y};\bar{x}_g,{\bf z}_1){\cal G}^{cb}(\bar{x}_g,{\bf z}_1;x_g,{\bf x}).
\end{align}
By applying this composition law to \eqref{aux1}, one arrives at two correlators encoding the color exchanges that take place within the light-cone time regions $(x_g,0)$ and $(L,x_g)$. Projecting them onto singlets is trivial in this case, as each has only two open color indices. For instance, the correlator of two `dressed' propagators becomes:
\begin{align}\label{dipole1}
\left\langle\left[{\cal G}{\cal G}^{\dagger}\right]^{ab}\right\rangle=\frac{\delta^{ab}}{N^2_c-1}\left\langle\text{Tr}\left\{{\cal G}{\cal G}^{\dagger}\right\}\right\rangle,
\end{align}
normalized to the dimension of the adjoint representation. By doing this, one finally obtains the following well-known result:
\begin{align}\label{BDMPSZ}
\langle|&{\cal M}|^2\rangle=\frac{g^2}{E_g^2}2\text{Re}\int^L_{0}\diff x_g\int^L_{x_g}\diff\bar{x}_g \, e^{i\frac{E_g}{2}{\bf n}^2(x_g-\bar{x}_g)}\int_{{\bf z}\,\bar{\bf z}\, {\bf z}_1} e^{-i{\bf k }\cdot({\bf z}- \bar{\bf z})}(i\partial_{{\bm \alpha}}\!-\!E_g{\bf n})(-i\partial_{{\bm \beta}}\!-\!E_g{\bf n})\nonumber\\
&\frac{1}{2N_c(N_c^2-1)}\left\langle \text{Tr}\left\{ {\cal G}(\bar{x}_g,{\bf z}_1;x_g,{\bm \alpha})U^{\dagger}(x_g,\bar{x}_g)\right\}\right\rangle\left\langle \text{Tr}\left\{ {\cal G}(L,\bar{\bf z};\bar{x}_g,{\bm \beta})\,{\cal G}(L,{\bf z};\bar{x}_g,{\bf z}_1)\right\}\right\rangle|_{\substack{\bm{\alpha}={\bf n}x_g\\ \bm{\beta}={\bf n}\bar{x}_g}},
\end{align}
which we will refer to as the BDMPS-Z spectrum.
This expression features two correlators: one involving a dressed propagator and an adjoint Wilson line, which describes an effective quark-gluon dipole; and another containing two dressed propagators, which encodes the momentum broadening of the emitted gluon. These are known, respectively, as the gluon emission kernel and the (coordinate-space representation of the) momentum broadening factor. We will use these correlators as a reference to assess the modifications introduced in our calculation. For further details on the derivation of the BDMPS-Z spectrum, see e.g.\ \cite{Casalderrey-Solana:2007knd,Casalderrey-Solana:2011ule}.

\section{Direct contribution to the in-medium $\gamma\rightarrow q\bar{q}g$ squared matrix element}
\label{sec:calculation}

\begin{figure}
\centering
\includegraphics[width=0.6\textwidth]{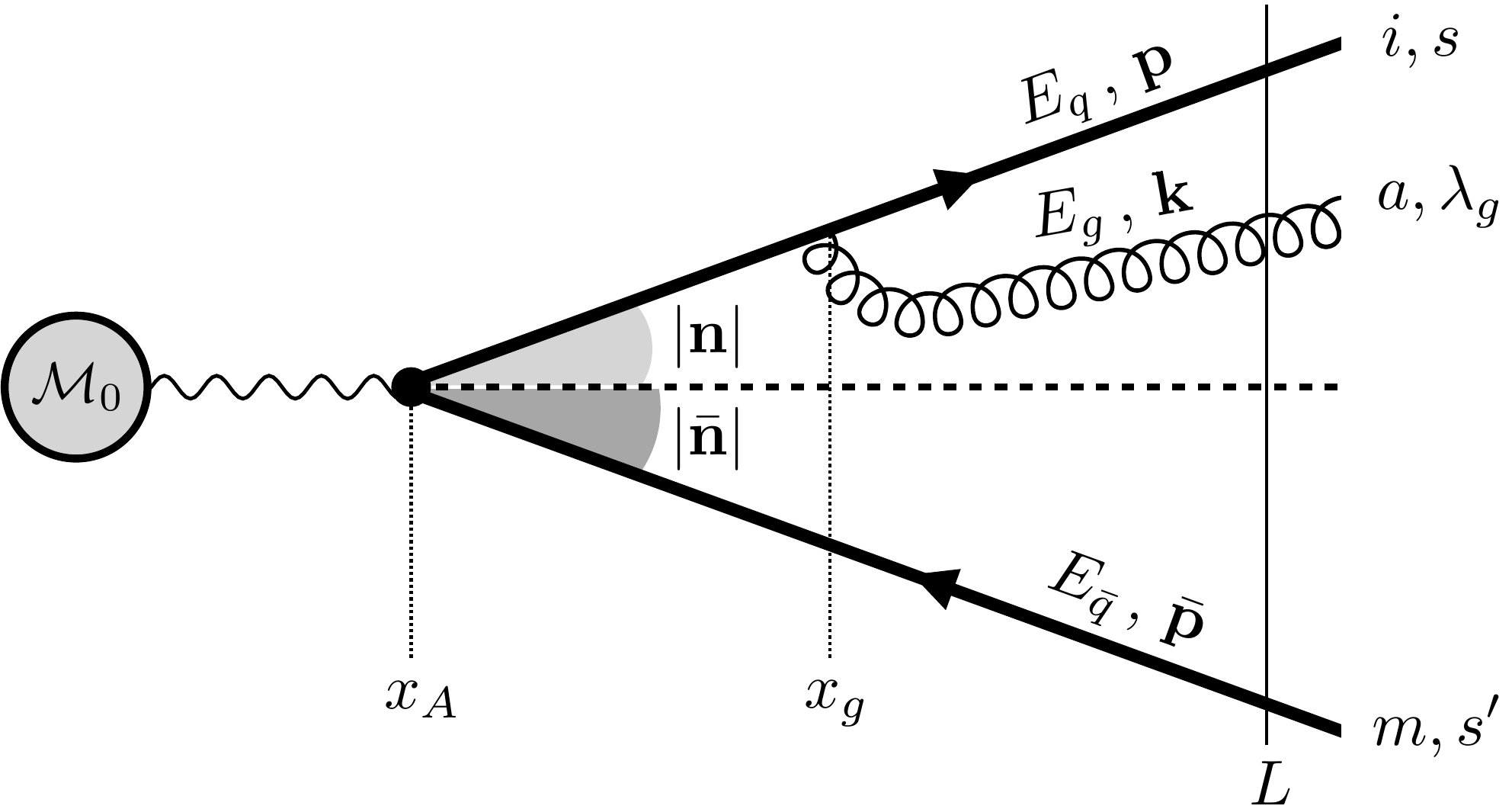}
\caption{Diagrammatic representation of the amplitude of a singlet quark-antiquark pair emitting a gluon within a dense QCD medium of length $L$.}
\label{fig:amplitude}
\end{figure}

We consider the emission of a soft gluon with light-cone energy $E_g$ and transverse momentum ${\bf k}$ off a color-singlet quark-antiquark antenna within a QGP of length $L$. The quark (antiquark) leg of the antenna carries energy $E_q$ ($E_{\bar{q}}$) and transverse momentum ${\bf p}$ $(\bar{\bf p})$, with the emitted gluon assumed to be soft in comparison. We focus on the case represented in \figref{fig:amplitude}, where the antenna formation and gluon emission respectively take place at light-cone times $x_A$ and $x_g$, which satisfy $x_A < x_g < L$ (i.e.\ the emission happens inside the medium). The corresponding amplitude reads:
\begin{align}\label{amp}
i{\cal M}^{im,ss';a,\lambda_g}_{\text{in}}({\bf k}, {\bf p},\bar{{\bf p}})\!= &\,-\frac{eg}{4E_{\gamma}(E_q+E_g)}e^{i\frac{{\bf k}^2}{2E_g}L}\!\!\!\int^L_0 \diff x_A \,e^{i\frac{x_A}{t_A}}\!\!\int^L_{x_A}\diff x_g\,e^{i\frac{E_g}{2}{\bf n}^2x_g} \!\!\int_{{\bf k}_1}\!e^{-i\,{\bf n}\cdot{\bf k}_1x_g}\!\!\int_{{\bf z},{\bf z}_1}\nonumber\\
&e^{-i({\bf k}\cdot{\bf z}-{\bf k}_1\cdot{\bf z}_1)}\,\Gamma^{ss_1}_{\lambda_g}[p,k_1]\,\Gamma^{s_1s'}_{\lambda_\gamma}[p,\bar{p}]\,{\cal M}^{\lambda_{\gamma}}_0(p+\bar{p})\nonumber\\
&{\cal G}^{ab}(L,{\bf z};x_g,{\bf z}_1)\left[V_1(L,x_g)t^bV_1(x_g,x_A)V^{\dagger}_2(x_A,L)\right]^{im},
\end{align}
where we have defined the formation time of the antenna as:
\begin{align}
t_A=\frac{2z_q(1-z_q)E_{\gamma}}{((1-z_q)({\bf p}+{\bf k}_1)+z_q\bar{{\bf p}})^2}\approx\frac{2z_q(1-z_q)E_{\gamma}}{((1-z_q){\bf p}+z_q\bar{{\bf p}})^2}.
\end{align}
Here, $E_{\gamma}\!=\!E_q+E_{\bar{q}}+E_g$ is the energy of the photon and the hermitian-conjugate Wilson line $V^{\dagger}$ describes the propagation of the antiquark. We adopt the shorthand notation $V^{kl}_i(\bar{t},t)\equiv V^{kl}(\bar{t},t;\left[{\bf r}_i\right])$ and define the following trajectories:
\begin{align}
{\bf r}_1(s)&\,\approx\frac{{\bf p}}{E_q}(s-x_A) \equiv {\bf n}(s-x_A)\label{qtraj}\\
{\bf r}_2(s)&\,=\frac{\bar{{\bf p}}}{E_{\bar{q}}}(s-x_A)\equiv\bar{\bf n}(s-x_A)
\end{align}
for quark and antiquark, respectively. Like we did in the previous section, we consider the quark to traverse the same trajectory (${\bf r}_1$) before and after gluon emission.
The photon-quark vertex is assumed to be unmodified by the presence of the medium, and reads:
\begin{align}
\Gamma^{ s_1s'}_{\lambda_\gamma}\left[p+k_1;\bar{p}\right]\approx\Gamma^{ s_1s'}_{\lambda_\gamma}\left[p;\bar{p}\right]\,&\!=\!\frac{2}{\sqrt{z_q(1-z_q)}}\delta_{-s_1s'}\left(z_q\delta_{\lambda_\gamma s'}-(1-z_q)\delta_{\lambda_\gamma s_1}\right)\!((1-z_q){\bf p}-z_q\bar{{\bf p}})\!\cdot\!\bm{\epsilon}_{\lambda_\gamma}\nonumber\\
&\equiv \gamma_{\lambda_\gamma}^{s_1s'}(z_q)((1-z_q){\bf p}-z_q\bar{{\bf p}})\!\cdot\!\bm{\epsilon}_{\lambda_\gamma},\label{eq:Yvertex}
\end{align}
where $z_q\!=\!(E_q+E_g)/E_\gamma$ is the quark's energy-sharing fraction. By substituting \eqref{eq:Yvertex} and \eqref{qgvertex} into \eqref{amp}, we arrive at:
\begin{align}\label{eq:third2}
i{\cal M}({\bf k}, {\bf p}, \bar{{\bf p}})=&\,-\frac{eg}{4E_{\gamma}(E_q+E_g)}\int^L_0 \diff x_A \, e^{i\frac{x_A}{t_A}}\int^L_{x_A}\diff x_g \, e^{i\frac{E_g}{2}{\bf n}^2x_g}\int_{{\bf z}} e^{-i{\bf k }\cdot {\bf z}} {\cal M}_0^{\lambda_\gamma}\gamma^{\lambda_g}_{ss_1}(z_g)\nonumber\\
&\hspace{-2cm}(i\partial_{{\bm \alpha}}\!-\!E_g{\bf n})\!\cdot\bm{\epsilon}^{*}_{\lambda_g}\gamma_{\lambda_\gamma}^{ss'}(z)\,{\bm P}\cdot\bm{\epsilon}_{\lambda_\gamma}{\cal G}^{ab}(L,{\bf z};x_g,{\bm \alpha})|_{\bm{\alpha}={\bf n}x_g}\left[V_1(L,x_g)t^bV_1(x_g,x_A)V^{\dagger}_2(x_A,L)\right]^{im}\!\!,
\end{align}
where we defined ${\bm P}\!\equiv\!(1-z_q){\bf p}-z_q\bar{\bf p}$ and neglected irrelevant phases.

Already at this stage, one can pinpoint several features that make the calculation of the squared amplitude substantially more involved than in the case examined in \secref{sec:setup}. While it is immediately apparent that \eqref{eq:third2} contains additional integrals, phases, and vertices compared to \eqref{eq:third1}, what is more relevant for our purposes is the presence of an extra Wilson line spanning the same light-cone time regions as the other propagators. Under the standard approximation of instantaneously-formed antennas, this Wilson line plays no role in the calculation, as it cancels exactly upon squaring the amplitude. However, as we will show below, it becomes crucial for the evaluation of the medium averages in our setup. Before turning to this discussion, we briefly outline the average (sum) over initial (final) quantum numbers involved in the squared amplitude. We start with the summation over the polarization of the photon, which yields: 
\begin{align}\label{average1}
\sum_{\lambda_\gamma \bar{\lambda}_\gamma}{\cal M}_0({\bf p})\cdot\epsilon^*_{\lambda_{\gamma}}\left({\cal M}^{\lambda_{\gamma}}_0(\bar{\bf p})\cdot\epsilon^*_{\lambda_{\gamma}}\right)^{\dagger}\rightarrow\sum_{\lambda_\gamma \bar{\lambda}_\gamma}\frac{\delta^{\lambda_\gamma \bar{\lambda}_\gamma}}{2}{\cal M}_0({\bf p}){\cal M}^{*}_0(\bar{\bf p})\equiv\sum_{\lambda_\gamma \bar{\lambda}_\gamma}\frac{\delta^{\lambda_\gamma \bar{\lambda}_\gamma}}{2}|{\cal M}_0|^2.
\end{align}
In this instance, $|{\cal M}_0|^2$ stands for the square of the virtual photon current, which will be omitted henceforth. This procedure also affects the product of vertices, which results in:
\begin{align}\label{average2}
\frac{1}{2}\sum\gamma_{\lambda_\gamma}^{ss'}\left(\gamma_{\lambda_\gamma}^{\bar{s}\bar{s}'}\right)^*\!\!=\frac{2}{z(1-z)}\left(z^2+(1-z)^2\right)=\frac{4E_{\gamma}}{t_A}\frac{1}{{\bf P}^2}\left(z^2+(1-z)^2\right).
\end{align}

\begin{figure}
\centering
\includegraphics[width=0.56\textwidth]{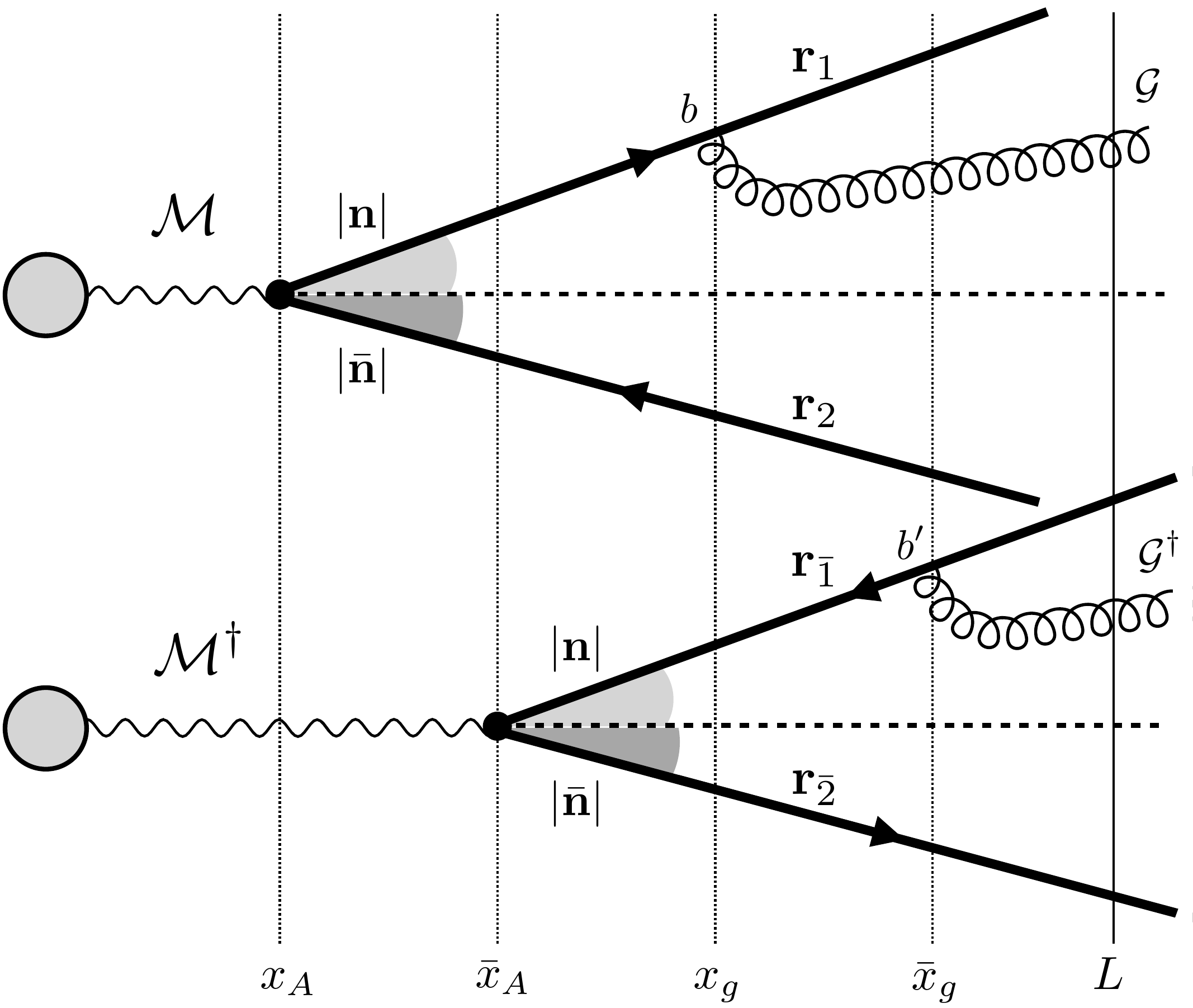}
\caption{Representation of the direct contribution to the squared amplitude of the in-medium $\gamma\rightarrow q\bar{q}g$ splitting.\vspace{-1em}}
\label{fig:radiation1}
\end{figure}

In this work we focus exclusively on the contributions where the gluon is emitted from the same leg in both amplitude and conjugate amplitude, i.e.\ the direct (or radiation) terms. We allow for interactions during the photon splitting by setting the formation of the antenna in the amplitude and conjugate amplitude at times $x_A$ and $\bar{x}_A$, respectively. The gluon emission takes place at time $x_g$ in the amplitude and $\bar{x}_g$ in the conjugate amplitude. We specifically consider the case depicted in \figref{fig:radiation1}, where $x_A<\bar{x}_A<x_g<\bar{x}_g<L$ and the gluon is emitted from the quark leg. Note, however, that our discussion would be identical if the roles of quark and antiquark were reversed.

By multiplying the corresponding amplitudes and substituting Eqs.\!~(\ref{average1}), (\ref{average2}), we arrive at:
\begin{align}\label{me1}
\hspace{-0.075cm}\langle|{\cal M}_q|^2\rangle =\, & {\cal M}_{q\bar{q}}^2 \left(\frac{2\,g}{E_g}\right)^{\!\!2}\int^L_0 \frac{\diff x_A}{t_A} \int^L_{x_A} \frac{\diff\bar{x}_A}{t_A} \, e^{i\frac{(x_A-\bar{x}_A)}{t_A}}\! \int^L_{\bar{x}_A}\diff x_g \int^L_{x_g}\diff\bar{x}_g \,e^{i\frac{E_g}{2}{\bf n}^2(x_g-\bar{x}_g)}\!\!\int_{{\bf z},\bar{{\bf z}}}\!\!e^{-i{\bf k }\cdot ({\bf z}-\bar{{\bf z}})}\nonumber\\
&(i\partial_{\bm{\alpha}}\!-\!E_g{\bf n})\!\cdot\! (-i\partial_{\bm{\beta}}\!-\!E_g{\bf n})\frac{1}{N_c}\Big\langle\big[{\cal G^{\dagger}}(\bar{x}_g,{\bm \beta};L,\bar{\bf z}){\cal G}(L,{\bf z};x_g,{\bm \alpha})\big]^{b'b}|_{\substack{\bm{\alpha}={\bf n}x_g\\ \bm{\beta}={\bf n}\bar{x}_g}}\nonumber\\
&\text{Tr}\big\{V^{\dagger}_{\bar{1}}(\bar{x}_A,\bar{x}_g)t^{b'}V^{\dagger}_{\bar{1}}(\bar{x}_g,L)V_1(L,x_g)t^bV_1(x_g,x_A)V^{\dagger}_2(x_A,L)V_{\bar{2}}(L,\bar{x}_A)\big\}\Big\rangle,
\end{align}
where we introduced the squared matrix element for the photon splitting as:
\begin{align}
{\cal M}_{q\bar{q}}^2=\frac{e^2t_A}{E_{\gamma}}P_{\gamma\rightarrow q\bar{q}}(z_q),
\end{align}
with $P_{\gamma\rightarrow q\bar{q}}=n_fN_c(z^2+(1-z)^2)$ the Altarelli-Parisi splitting function. Here we have also introduced the following Wilson line paths:
\begin{align}
{\bf r}_{\bar{1}}(s)&\,\approx{\bf n}(s-\bar{x}_A)\\
{\bf r}_{\bar{2}}(s)&\,=\bar{\bf n}(s-\bar{x}_A),
\end{align}
corresponding respectively to the trajectories traversed by quark and antiquark in the conjugate amplitude.

For an instantaneously-formed antenna, instead of treating $x_A$, $\bar{x}_A$ as integration variables, one simply sets $x_A\!=\!\bar{x}_A\!=\!t_0$, with fixed $t_0$ (which may be set to $0$).
This implies ${\bf r}_2\!=\!{\bf r}_{\bar{2}}$, leading to the mutual cancelation of the antiquark lines in amplitude and conjugate amplitude. This limit also implies ${\bf r}_1\!=\!{\bf r}_{\bar{1}}$, resulting in a partial cancelation of the remaining Wilson lines, which can then be recast as a single adjoint Wilson line. The corresponding squared amplitude thus becomes identical (up to a pre-factor) to the one computed in \secref{sec:setup}.

In the present setup, however, the Wilson line paths are separated by transverse `gaps' proportional to $\bar{x}_A-x_A\equiv\Delta x_A$, preventing these cancelations. As a consequence, more than twice as many propagators enter the medium average, which severely complicates the required color algebra manipulations. Specifically, these manipulations entail decomposing the medium average in \eqref{me1} (denoted as $\langle...\rangle$ henceforth) into correlators of color-singlet states.
This task was straightforward in the case examined in \secref{sec:setup}, where each time region effectively contained only two particles. The only way for such a system to exchange color with the medium without altering its overall color state (as enforced by the Gaussian ansatz) is in a dipole configuration.
Nevertheless, since we now deal with two or more particles per time region, the number of possible combinations increases accordingly. To account for all of them, one must project onto bases whose dimension ranges from 2 (in the simplest case) to 6 (in the most involved one), leading to lengthy algebraic manipulations. In order to streamline the main body of the manuscript, we outline these calculations in a set of appendices (\ref{app:medav} and \ref{app:adjoint}), divided into two sections for clarity. The final result of the procedure, written in matrix notation, is:
\begin{align}\label{finalrad}
\langle...\rangle={\cal C}\!\int_{{\bf z}_1} \!\left\langle\text{Tr}\left\{V_1V^{\dagger}_2\right\}\right\rangle_{(\bar{x}_A,x_A)} \left\langle {\bold A}\right\rangle_{(x_g,\bar{x}_A)} \!\cdot{\bold g}_2\cdot\left\langle {\cal G}^{ab}(\bar{x}_g,{\bf z}_1;x_g,{\bm \alpha}) \, {\bold B}^{ba}\right\rangle_{(\bar{x}_g,x_g)}\nonumber\\
\cdot\,{\bold g}_4\!\cdot\!\left\langle {\cal G}^{cd}(L,{\bf z};\bar{x}_g,{\bf z}_1)\,{\bold C}^{de}\, {\cal G}^{\dagger ec}(\bar{x}_g,{\bm \beta};L,\bar{\bf z})\right\rangle_{(L,\bar{x}_g)},
\end{align}
where we defined the color pre-factor ${\cal C}\!=\!\frac{2}{N^3_c(N_c^2-1)^2(N_c^2-4)}$.
Note that, in order to arrive at this expression, we needed to split a gluon propagator into regions, which brings an integration over the intermediate transverse position ${\bf z}_1$ (see \eqref{Gsplit}). The vectors ${\bold A}$, ${\bold B}$, ${\bold C}$ contain all the allowed combinations of Wilson lines within traces for the $(x_g,\bar{x}_A)$, $(\bar{x}_g,x_g)$ and $(L,\bar{x}_g)$ time regions, respectively. Since the former region contains four Wilson lines, the corresponding vector has only two elements:
\begin{align}\label{WLvecsA}
{\bold A}\!=\!\left(\text{Tr}\{ V_1 V^{\dagger}_2\}\text{Tr}\{ V_{\bar{2}} V^{\dagger}_{\bar{1}}\},\,\text{Tr}\{ V_1 V^{\dagger}_2V_{\bar{2}} V^{\dagger}_{\bar{1}}\}\right).
\end{align}
However, the color algebra in the latter regions is significantly more involved, giving rise to the following combinations:\vspace{-0.5em}
\begin{align}\label{WLvecs}
\left({\bold B}^{ba}\right)^{\text{T}}\!\!\!\!=\!\!\begin{pmatrix} \text{Tr}\{ V_1 t^b V^{\dagger}_2V_{\bar{2}} V^{\dagger}_{\bar{1}}t^a\} \!\!&\!\!\text{Tr}\{ t^bV^{\dagger}_{\bar{1}} t^a V_1 \}\text{Tr}\{ V_{\bar{2}} V^{\dagger}_{2} \} \\   \text{Tr}\{ V_1 t^bV^{\dagger}_2 \}\text{Tr}\{V^{\dagger}_{\bar{1}}t^aV_{\bar{2}} \} \!\!&\!\! \text{Tr}\{ t^b V^{\dagger}_{\bar{1}} t^a V_{\bar{2}} V^{\dagger}_{2}V_1\}\\  \text{Tr}\{ t^bV^{\dagger}_2 t^a V_1 \}\text{Tr}\{ V_{\bar{2}} V^{\dagger}_{\bar{1}} \} \!\!&\!\!\text{Tr}\{ t^b V^{\dagger}_{\bar{1}}V_{\bar{2}} V^{\dagger}_{2}t^a V_1\}\\  \text{Tr}\{ t^b V^{\dagger}_{2} t^a V_{\bar{2}} V^{\dagger}_{\bar{1}}V_1 \} \!\!&\!\!  \text{Tr}\{ V_1 t^bV^{\dagger}_{\bar{1}} \}\text{Tr}\{V^{\dagger}_{2}t^aV_{\bar{2}} \} \end{pmatrix}\!,{\bold C}^{de} \!=\!\!\begin{pmatrix} \text{Tr}\{  t^dt^e V^{\dagger}_{\bar{1}} V_1 \}\text{Tr}\{ V^{\dagger}_{2}V_{\bar{2}}  \}  \\  \text{Tr}\{ t^d t^e V^{\dagger}_{\bar{1}} V_1 V^{\dagger}_{2} V_{\bar{2}}  \}  \\  \text{Tr}\{t^d V^{\dagger}_{2}V_{\bar{2}}  t^e V^{\dagger}_{\bar{1}} V_1  \}  \\ \text{Tr}\{ V_{\bar{2}}t^dV^{\dagger}_{2}  \} \text{Tr}\{  V_1t^e V^{\dagger}_{\bar{1}}\}\end{pmatrix}\!.\\[-3em]\nonumber
\end{align}
The products between these vectors are mediated by the following matrices:
\begin{align}\label{metric}
{\bold g}_2 =\begin{pmatrix}
N_c & -1 \\
-1 & N_c
\end{pmatrix}\!,\,{\bold g}_4=&\,\begin{pmatrix}
(N_c^2-2) & -N_c & -N_c & 2 \\
-N_c & (N_c^2-2)  & 2 & -N_c \\
-N_c & 2 & (N_c^2-2) & -N_c \\
 2 & -N_c & -N_c & (N_c^2-2)
\end{pmatrix},
\end{align}
which, along with ${\cal C}$, encode the explicit $N_c$-dependence of \eqref{finalrad}.

This result differs from the BDMPS-Z spectrum in several key ways. Most notably, it features a large number of independent contributions (160 when written in the fundamental representation), each corresponding to a distinct way in which the system can evolve while preserving its total color charge. Another crucial difference lies in the internal structure of these contributions. While the BDMPS-Z spectrum involves two correlators (encoding the medium interactions that take place during and after gluon emission), each term in \eqref{finalrad} contains four correlators instead, corresponding to the four regions shown in \figref{fig:radiation1}. The first one, common to all terms, is the Wilson line dipole describing the decoherence experienced by the quark-antiquark pair during its finite formation time: $\langle\text{Tr}\{V_1V^{\dagger}_2\}\rangle_{(\bar{x}_A,x_A)}$. For the second correlator,
two distinct possibilities branch out: either a double dipole or a quadrupole (see \eqref{WLvecsA}).
The remaining two correlators encode, respectively, the emission and subsequent broadening of a gluon, thus being analogous to those featured in the spectrum off a single charge. However, the larger number of particles involved in the corresponding correlators increases the number of available color states, which is reflected in the dimensions of ${\bold A}$, ${\bold B}$, and ${\bold C}$, as shown in \eqref{WLvecs}.

As a consistency check, we verify that the characteristic correlators from the BDMPS-Z spectrum are recovered in the limit where the antenna forms instantaneously. Since we are focusing our analysis on the medium average, this limit is adopted simply by setting $x_A=\bar{x}_A$, which in turn makes the Wilson line trajectories appearing in amplitude and conjugate amplitude identical: ${\bf r}_1\!=\!{\bf r}_{\bar{1}}$, ${\bf r}_2\!=\!{\bf r}_{\bar{2}}$. By doing this and expanding \eqref{finalrad}, we eventually arrive at a single term:
\begin{align}\label{limit1}
\langle...\rangle\!\underset{x_A=\,\bar{x}_A}{\longrightarrow}\frac{1}{2N_c(N_c^2-1)}\int_{{\bf z}_1}&\left\langle \text{Tr}\left\{{\cal G}(\bar{x}_g,{\bf z}_1;x_g,\bm{\alpha})U^{\dagger}_{1}\right\} \right\rangle_{(\bar{x}_g,x_g)}\nonumber\\
&\times\!\left\langle \text{Tr}\left\{{\cal G}(L,{\bf z};\bar{x}_g,{\bf z}_1){\cal G}^{\dagger}(\bar{x}_g,\bm{\beta};L,\bar{{\bf z}})\right\} \right\rangle_{(L,\bar{x}_g)}.
\end{align}
Here we identify the gluon emission kernel (first correlator) and gluon-gluon dipole (second correlator), as well as the correct color factor.

An additional check consists in assuming that the medium vanishes at a light-cone time $L'\!<\!x_g$, i.e.\ before the gluon emission takes place. The purpose of this exercise is to verify that both the overall $N_c$-scaling and the specific correlators emerging in this scenario are consistent with previous results. To this end, we once more restrict the analysis to the level of the medium average, ignoring other parts of the calculation such as the integration over the intermediate position ${\bf z}_1$. Therefore, the procedure simply amounts to replacing all Wilson lines of regions $(\bar{x}_g,x_g)$ and $(L,\bar{x}_g)$ with the identity. By doing so, we obtain:
\begin{align}
\langle...\rangle\rightarrow C_FN_c \left\langle S_{12}\right\rangle_{(\bar{x}_A,x_A)}\left\langle Q_{12\bar{2}\bar{1}} \right\rangle_{(L',\bar{x}_A)},\label{eq:limit2}
\end{align}
where we adopted the following notation for the dipole and quadrupole functions:
\begin{align}
S_{ij}=&\,\frac{1}{N_c}\text{Tr}\left\{V_iV^{\dagger}_j\right\}\label{dip}\\
Q_{ijkl}=&\,\frac{1}{N_c}\text{Tr}\left\{V_iV^{\dagger}_jV_kV^{\dagger}_l\right\}.\label{quad}
\end{align}
\eqref{eq:limit2} is in agreement with the analogous result obtained in \cite{Abreu:2024wka}.

\subsection{Leading-$N_c$ contribution}
\label{sec:largenc}

\begin{figure}
\centering
\makebox[\textwidth][c]{\includegraphics[width=0.84\textwidth]{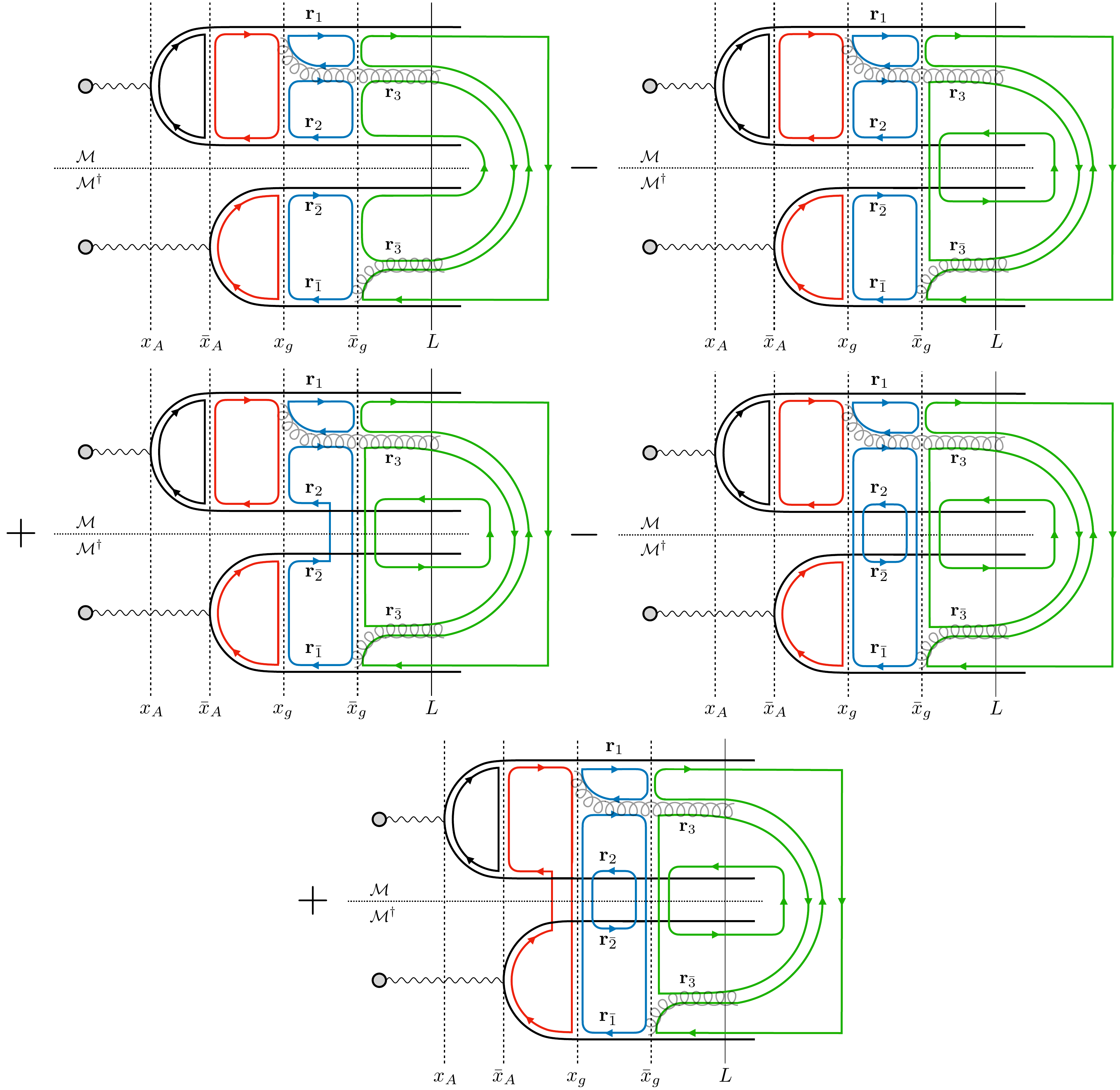}}
\caption{Diagrammatic representation of the correlators featured in each term of the leading-$N_c$ contribution, \eqref{largenc1}, drawn in the same order as they are written. For the sake of visual clarity, in these diagrams the quark/antiquark positions are mirrored with respect to the separation between amplitude and conjugate amplitude.}
\label{fig:color1}
\end{figure}

As mentioned above, the matrices ${\bold g}_2$, ${\bold g}_4$ encode the explicit $N_c$-power hierarchy of the terms contained in \eqref{finalrad}. However, in order to identify the dominant contribution in the large-$N_c$ limit, one must also account for the {\it implicit} $N_c$-dependence of the Wilson line correlators. In the fundamental representation, this can be done straightforwardly by counting the number of traces they contain\footnote{In the large-$N_c$ limit, the correlator $\langle\text{Tr}_1\{V^{\dagger}V...V^{\dagger}V\}...\text{Tr}_k\{V^{\dagger}V...V^{\dagger}V\}\rangle$ simplifies to $\langle\text{Tr}_1\{V^{\dagger}V...V^{\dagger}V\}\rangle...\langle\text{Tr}_k\{V^{\dagger}V...V^{\dagger}V\}\rangle$, which scales like $N_c^k$.}. This reasoning, however, cannot be directly applied to \eqref{finalrad}, since it involves adjoint Wilson lines. Instead, the $N_c$ power counting is most naturally carried out at one of the intermediate stages of our calculation, shown in appendix~\ref{app:medav}, where all Wilson lines appear in the fundamental representation. Specifically, the total $N_c$-scaling of any given term in \eqref{fundrep} is determined simply by adding the number of traces it contains to the explicit power of $N_c$ that it carries.
This procedure yields the following five terms:
\begin{align}\label{largenc1}
\langle...\rangle\!\underset{N_c\uparrow}{\rightarrow}\!{\cal C}_f\!\!\int_{{\bf z}_1}\!\int {\cal D} {\bf r}_3 \,{\cal D} {\bf r}_{\bar{3}} \,\left\langle S_{12} \right\rangle_{\scriptscriptstyle (\bar{x}_A,x_A)}\!\!\Bigg[\left\langle S_{12} \, S_{\bar{2}\bar{1}}\right. &\left. \right\rangle_{\scriptscriptstyle (x_g,\bar{x}_A)}\!
\bigg(\!\!\left\langle S_{13}\,S_{32}\,S_{\bar{2}\bar{1}}\right\rangle_{\scriptscriptstyle (\bar{x}_g,x_g)}\left\langle Q_{\bar{2}\,\bar{3}\,3\,2}\,Q_{\bar{3}\,\bar{1}\,1\,3}\right\rangle_{\scriptscriptstyle (L,\bar{x}_g)} \nonumber\\
&-\left\langle S_{13}\,S_{32}\,S_{\bar{2}\bar{1}}\right\rangle_{\scriptscriptstyle (\bar{x}_g,x_g)}\left\langle S_{\bar{2}2}\,S_{3\bar{3}}\,Q_{\bar{3}\,\bar{1}\,1\,3}\right\rangle_{\scriptscriptstyle (L,\bar{x}_g)} \nonumber\\
&+\left\langle S_{13}\, Q_{3\,2\,\bar{2}\,\bar{1}}\right\rangle_{\scriptscriptstyle (\bar{x}_g,x_g)}\!\left\langle S_{\bar{2}2}\,S_{3\bar{3}}\,Q_{\bar{3}\,\bar{1}\,1\,3}\right\rangle_{\scriptscriptstyle (L,\bar{x}_g)} \nonumber\\
&-\left\langle S_{13}\,S_{3\bar{1}}\,S_{\bar{2}2}\right\rangle_{\scriptscriptstyle (\bar{x}_g,x_g)}\left\langle S_{\bar{2}2}\,S_{3\bar{3}}\,Q_{\bar{3}\,\bar{1}\,1\,3}\right\rangle_{\scriptscriptstyle (L,\bar{x}_g)}\bigg) \nonumber\\
&\hspace{-4.cm}+\left\langle Q_{1\,2\,\bar{2}\,\bar{1}}\right\rangle_{\scriptscriptstyle (x_g,\bar{x}_A)}\bigg(\left\langle S_{13}\,S_{3\bar{1}}\,S_{\bar{2}2}\right\rangle_{\scriptscriptstyle (\bar{x}_g,x_g)}
\left\langle S_{\bar{2}2}\,S_{3\bar{3}}\,Q_{\bar{3}\,\bar{1}\,1\,3}\right\rangle_{\scriptscriptstyle (L,\bar{x}_g)}\bigg)\Bigg],
\end{align}
where we defined the overall color factor as ${\cal C}_f=\frac{N_c^8}{2(N_c^2-1)^2(N_c^2-4)}$.
Note that, in the large-$N_c$ limit, this factor scales like $N_c^2/2\approx C_FN_c$.
Apart from the quark (antiquark) trajectories in the amplitude and conjugate amplitude, denoted respectively by ${\bf r}_{1(2)}$ and ${\bf r}_{\bar{1}(\bar{2})}$, this expression also features the gluon paths ${\bf r}_3$ and ${\bf r}_{\bar{3}}$. Since these trajectories are associated with `dressed' propagators, they are integrated over in path integrals, as explicitly indicated in \eqref{largenc1}. For simplicity, however, we omit the kinematic exponential factors also carried by these propagators. The color flow encoded in these terms is represented diagrammatically in \figref{fig:color1}.

The physical interpretation of \eqref{largenc1} becomes clearer after we use the Fierz identity,
\begin{align}
t^a_{ij}t^a_{kl}\!=\!\frac{1}{2}\!\left(\delta_{il}\delta_{jk}-\frac{1}{N_c}\delta_{ij}\delta_{kl}\right)\!,\label{eq:fierz}
\end{align}
to insert products of color matrices into our expression. This allows us to recombine the first four terms by pairs.
For instance, the first two terms can be written as:
\begin{align}\label{sublead}
\left\langle S_{12}\right\rangle_{(\bar{x}_A,x_A)}\!\left\langle S_{12}S_{\bar{2}\bar{1}}\right\rangle_{(x_g,\bar{x}_A)}\bigg[\!\left\langle S_{13}S_{32}S_{\bar{2}\bar{1}}\right\rangle_{(\bar{x}_g,x_g)}\!\bigg(\left\langle Q_{\bar{2}\bar{3}32}Q_{\bar{3}\bar{1}13}\right\rangle-\left\langle S_{\bar{2}2}S_{3\bar{3}}Q_{\bar{3}\bar{1}13}\right\rangle\bigg)_{\!(L,\bar{x}_g)}\bigg]=\nonumber\\
\frac{2}{N_c}\left\langle S_{12}\right\rangle_{(\bar{x}_A,x_A)}\!\left\langle S_{12}S_{\bar{2}\bar{1}}\right\rangle_{(x_g,\bar{x}_A)}\bigg[\!\left\langle S_{13}S_{32}S_{\bar{2}\bar{1}}\right\rangle_{(\bar{x}_g,x_g)}\left\langle \text{Tr}\{ V_{\bar{2}}t^aV^{\dagger}_2\}\text{Tr}\{ V_{3}t^aV^{\dagger}_{\bar{3}}\}Q_{\bar{3}\bar{1}13}\right\rangle_{\!(L,\bar{x}_g)}\bigg],
\end{align}
and similarly for the following two terms. Likewise, one can expand the quadrupole featured in region $(x_g,\bar{x}_A)$ of the last term of \eqref{largenc1} as:
\begin{align}\label{quad1}
Q_{12\bar{2}\bar{1}}=S_{1\bar{1}}S_{\bar{2}2}+\frac{2}{N_c}\text{Tr}\{ V_1t^aV^{\dagger}_{\bar{1}}\}\text{Tr}\{ V_{\bar{2}}t^aV^{\dagger}_{2}\}.
\end{align}
Upon performing the medium average and taking the large-$N_c$ limit, this expression becomes analogous to the well-known decomposition of a Wilson line quadrupole into factorizable and non-factorizable pieces, first introduced in \cite{Blaizot:2012fh} (see appendix~\ref{app:limit} for details).

\begin{figure}
\centering
\makebox[\textwidth][c]{\includegraphics[width=1.29\textwidth]{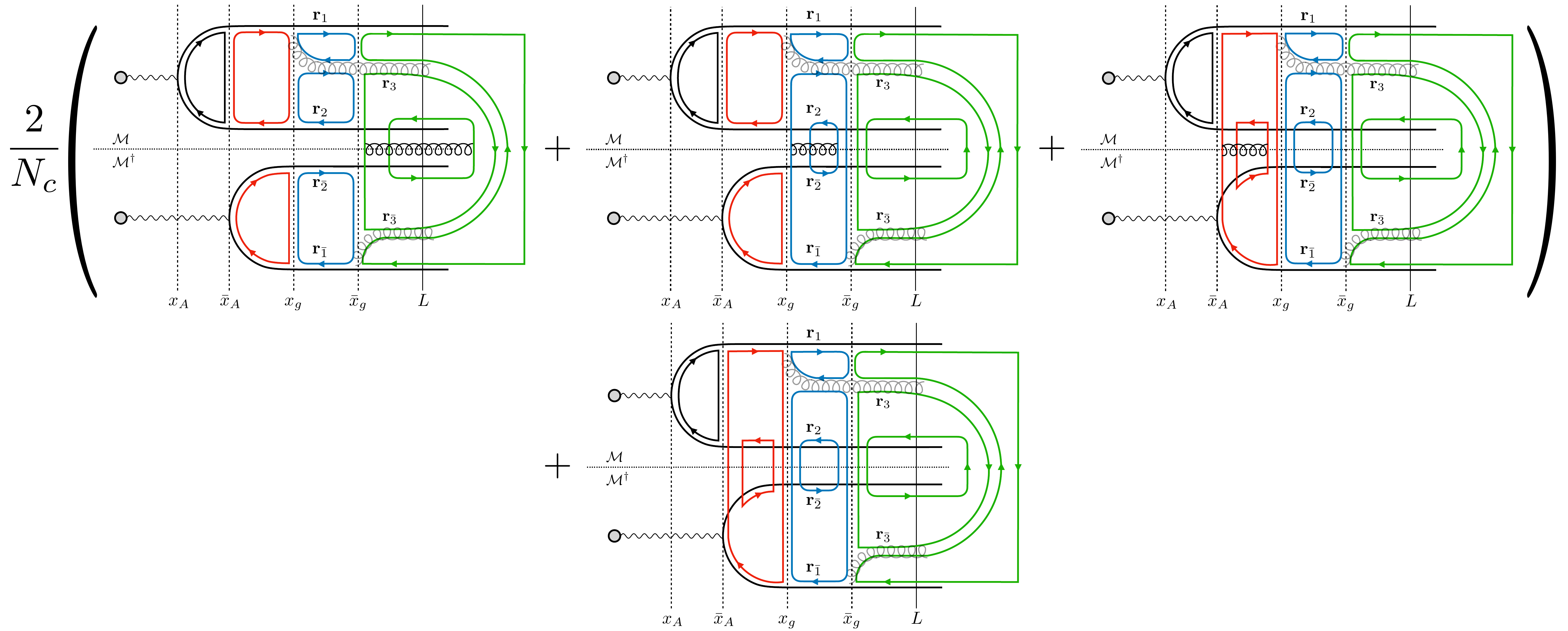}}
\caption{Diagrammatic representation of \eqref{largenc1b}. The presence of contracted color generators is denoted by a thick gluon line connecting color loops.}
\label{fig:color1new}
\end{figure}

By applying these transformations, \eqref{largenc1} becomes:
\begin{align}\label{largenc1b}
\langle...\rangle\!\underset{N_c\uparrow}{\rightarrow}&\,\,{\cal C}_f\!\!\int_{{\bf z}_1}\!\int {\cal D} {\bf r}_3 \,{\cal D} {\bf r}_{\bar{3}} \,\langle S_{12}\rangle_{\scriptscriptstyle (\bar{x}_A,x_A)}\bigg[\nonumber\\
&\,\,\,\,\,\,\frac{2}{N_c}\left\langle S_{12} \, S_{\bar{2}\bar{1}}\right\rangle_{\scriptscriptstyle (x_g,\bar{x}_A)}\left\langle S_{13}\,S_{32}\,S_{\bar{2}\bar{1}}\right\rangle_{\scriptscriptstyle (\bar{x}_g,x_g)}\!\!\left\langle \text{Tr}\{ V_{\bar{2}}t^aV^{\dagger}_2\}\text{Tr}\{V_3t^a V^{\dagger}_{\bar{3}}\}\,Q_{\bar{3}\,\bar{1}\,1\,3}\right\rangle_{\scriptscriptstyle (L,\bar{x}_g)} \nonumber\\
&+\frac{2}{N_c}\left\langle S_{12} \, S_{\bar{2}\bar{1}}\right\rangle_{\scriptscriptstyle (x_g,\bar{x}_A)}\left\langle \text{Tr}\{ V_{3}t^aV^{\dagger}_{\bar{1}}\}\text{Tr}\{V_{\bar{2}}t^a V^{\dagger}_{2}\}\,S_{13}\right\rangle_{\scriptscriptstyle (\bar{x}_g,x_g)}\!\!\left\langle S_{\bar{2}2}S_{3\bar{3}}\,Q_{\bar{3}\,\bar{1}\,1\,3}\right\rangle_{\scriptscriptstyle (L,\bar{x}_g)} \nonumber\\
&+\frac{2}{N_c}\left\langle \text{Tr}\{ V_1t^aV^{\dagger}_{\bar{1}}\}\text{Tr}\{ V_{\bar{2}}t^aV^{\dagger}_{2}\}\right\rangle_{\scriptscriptstyle (x_g,\bar{x}_A)}\left\langle S_{13}S_{3\bar{1}}\,S_{\bar{2}2}\right\rangle_{\scriptscriptstyle (\bar{x}_g,x_g)}\!\!\left\langle S_{\bar{2}2}S_{3\bar{3}}\,Q_{\bar{3}\,\bar{1}\,1\,3}\right\rangle_{\scriptscriptstyle (L,\bar{x}_g)} \nonumber\\
&+\left\langle S_{1\bar{1}} \, S_{\bar{2}2}\right\rangle_{\scriptscriptstyle (x_g,\bar{x}_A)}\left\langle S_{13}S_{3\bar{1}}\,S_{\bar{2}2}\right\rangle_{\scriptscriptstyle (\bar{x}_g,x_g)}\!\!\left\langle S_{\bar{2}2}S_{3\bar{3}}\,Q_{\bar{3}\,\bar{1}\,1\,3}\right\rangle_{\scriptscriptstyle (L,\bar{x}_g)}\bigg].
\end{align}
This result is represented in \figref{fig:color1new}, where we depict the contracted color generators as an effective gluon exchange between dipoles\footnote{Note, however, that these contractions encode the exchange of an arbitrary amount of gluons, reducing to a single gluon exchange only in the large-$N_c$ limit (see appendix~\ref{app:limit}).}. 
Besides being slightly more compact, the above expression makes the physical interpretation of each contribution significantly clearer. For instance, the last term encodes a color exchange in which, except during the antenna formation period $(\bar{x}_A,x_A)$, the antiquark lines in the amplitude and conjugate amplitude evolve in a dipole configuration, i.e.\ $S_{\bar{2}2}$. In the large-$N_c$ limit (where the correlators factorize), this corresponds to an antiquark rotating its color independently of quark and gluon throughout the interaction. This contribution therefore amounts to the BDMPS-Z spectrum multiplied by a simple correction factor.

The remaining three terms (which we call {\it transition} terms) do contain explicit correlations with the antiquark, even in the large-$N_c$ limit. These appear in the form of dipoles (such as $S_{12}$ or $S_{32}$) and effective gluon exchanges. In the case of the first term, the exchange takes place between the antiquark and the gluon: $\text{Tr}\{ V_{\bar{2}}t^aV^{\dagger}_2\}\text{Tr}\{V_3t^a V^{\dagger}_{\bar{3}}\}$, within the latest light-cone time region. In the second term, it happens within the region corresponding to the gluon formation, $(\bar{x}_g,x_g)$. As quark and gluon are unresolved during this period, the antiquark interacts with a dipole state that includes Wilson lines from both particles: $\text{Tr}\{ V_{3}t^aV^{\dagger}_{\bar{1}}\}\text{Tr}\{V_{\bar{2}}t^a V^{\dagger}_{2}\}$. Finally, the third term encodes an effective gluon exchange between the quark and the antiquark, which takes place within the $(x_g,\bar{x}_A)$ region. In all cases, such exchanges signal the onset of configurations in which, in the large-$N_c$ limit, the antiquark becomes decorrelated at subsequent light-cone times.

Due to color screening, the transition terms are expected to be suppressed when the antiquark is well separated from the other particles. This assumption is valid when the opening angle of the antenna is large and the gluon is emitted a long time after its formation. To see this explicitly, we can apply the large-$N_c$ limit in combination with the harmonic approximation to compute certain correlators, for instance:
\begin{align}
\left\langle S_{12}\right\rangle_{(x_g,\bar{x}_A)}=\exp\left\{ -\frac{\hat{q}}{12} \theta_{q\bar{q}}^2\left((x_g-x_A)^3-\Delta x_A^3\right)\right\}.
\end{align}
The expression above shows that large values of the antenna opening angle (defined as $|{\bf n}-\bar{\bf n}|\equiv\theta_{q\bar{q}}$) lead to an exponential suppression of the first two terms in \eqref{largenc1b} with respect to the last one.
To demonstrate that the third term is also negligible in this regime (although to a lesser extent), one needs to show that a similar exponential suppression arises from the effective gluon exchanges as well. This is done in appendix~\ref{app:limit}.

Finally, it is possible to combine \eqref{largenc1} with $N_c$-subleading terms in order to recover the adjoint Wilson lines featured in our main result, thereby allowing for a more direct comparison to the BDMPS-Z spectrum. This procedure is explicitly illustrated in appendix~\ref{app:adjoint} using matrix notation.
We arrive at:
\begin{align}\label{largenc}
\langle...\rangle\approx {\cal C}_a\int_{{\bf z}_1}\hspace{13.cm}\nonumber\\
\left\langle S_{12}\right\rangle_{\!\scriptscriptstyle (\bar{x}_A,x_A)}\!\!\Bigg[\!\!\left\langle S_{12}\,S_{\bar{2}\bar{1}}\right\rangle_{\scriptscriptstyle (x_g,\bar{x}_A)}\!\bigg(\!\!\left\langle{\cal G}^{ab}\text{Tr}\{ t^bV^{\dagger}_2 t^a V_1 \}S_{\bar{2}\bar{1}}\right\rangle_{\!\!\scriptscriptstyle (\bar{x}_g,x_g)}
\!\!\left\langle{\cal G}^{cd}\text{Tr}\{  t^d V^{\dagger}_{2}V_{\bar{2}}t^e V^{\dagger}_{\bar{1}} V_1   \}{\cal G}^{\dagger ec}\right\rangle_{\!\!\scriptscriptstyle (L,\bar{x}_g)}\nonumber\\
-\left\langle{\cal G}^{ab}\text{Tr}\{ t^bV^{\dagger}_2 t^a V_1 \}S_{\bar{2}\bar{1}}\right\rangle_{\!\!\scriptscriptstyle (\bar{x}_g,x_g)}
\!\!\left\langle{\cal G}^{cd}\text{Tr}\{  t^dt^e V^{\dagger}_{\bar{1}} V_1 \}S_{\bar{2}2}{\cal G}^{\dagger ec}\right\rangle_{\!\!\scriptscriptstyle (L,\bar{x}_g)}\nonumber\\
+\left\langle{\cal G}^{ab}\text{Tr}\{ V_1t^b V^{\dagger}_2V_{\bar{2}} V^{\dagger}_{\bar{1}}t^a \}\right\rangle_{\!\! \scriptscriptstyle (\bar{x}_g,x_g)}
\!\!\left\langle{\cal G}^{cd}\text{Tr}\{  t^dt^e V^{\dagger}_{\bar{1}} V_1 \}S_{\bar{2}2}{\cal G}^{\dagger ec}\right\rangle_{\!\! \scriptscriptstyle(L,\bar{x}_g)}\nonumber\\
-\left\langle{\cal G}^{ab}\text{Tr}\{ t^b V^{\dagger}_{\bar{1}} t^a V_1 \}S_{\bar{2}2}\right\rangle_{\!\!\scriptscriptstyle (\bar{x}_g,x_g)}
\!\!\left\langle{\cal G}^{cd}\text{Tr}\{  t^dt^e V^{\dagger}_{\bar{1}} V_1 \}S_{\bar{2}2}{\cal G}^{\dagger ec}\right\rangle_{\!\!\scriptscriptstyle (L,\bar{x}_g)}\!\!\bigg)\nonumber\\
+\left\langle Q_{12\bar{2}\bar{1}}\right\rangle_{\scriptscriptstyle (x_g,\bar{x}_A)}\!\!\bigg(\!\!\left\langle{\cal G}^{ab}\text{Tr}\{ t^b V^{\dagger}_{\bar{1}} t^a V_1 \}S_{\bar{2}2}\right\rangle_{\!\!\scriptscriptstyle (\bar{x}_g,x_g)}
\!\!\left\langle{\cal G}^{cd}\text{Tr}\{  t^dt^e V^{\dagger}_{\bar{1}} V_1\}S_{\bar{2}2}{\cal G}^{\dagger ec}\right\rangle_{\!\!\scriptscriptstyle (L,\bar{x}_g)}\bigg)\Bigg],
\end{align}
which carries the overall color factor ${\cal C}_a\!=\!\frac{2N_c^4}{(N_c^2-1)^2(N_c^2-4)}$. We remark that, although the above expression does include some contributions that are absent in \eqref{largenc1}, both results are equivalent at large $N_c$. This may be verified by expanding the correlators from regions $(\bar{x}_g,x_g)$ and $(L,\bar{x}_g)$ by means of the Fierz identity. For instance, let us do this for the $(\bar{x}_g,x_g)$-region correlator of the third term in \eqref{largenc}. By focusing on the Wilson line component of the `dressed' propagator (which is the only one relevant to the medium average), we have:
\begin{align}
\left\langle
U^{ab}_3\text{Tr}\{ V_1t^b V^{\dagger}_2V_{\bar{2}} V^{\dagger}_{\bar{1}}t^a\}
\right\rangle=\left\langle
\left[V^{\dagger}_3t^aV_3\right]_{ij}\left[V^{\dagger}_2V_{\bar{2}}V^{\dagger}_{\bar{1}}t^aV_1\right]_{ji}
\right\rangle\underset{N_c\uparrow}{\rightarrow}\frac{N_c^2}{2}\left\langle S_{13} \,Q_{32\bar{2}\bar{1}}\right\rangle.
\end{align}
In the first equality we used the identity $U^{ab}t^b_{ij}\!=\!\left[V^{\dagger}t^aV\right]_{ij}$ to transform the adjoint Wilson line to fundamental representation. Then, we applied the Fierz identity and neglected the $N_c$-suppressed term, thus arriving at the final result. By acting similarly upon the $(L,\bar{x}_g)$-region correlator one obtains the third term of \eqref{largenc1}. Note that this procedure (iterated for every correlator in \eqref{largenc} containing a `dressed' propagator) also accounts for the overall factor $N_c^4/4$ difference between ${\cal C}_f$ and ${\cal C}_a$.

Given these expressions, it is worthwhile to repeat the checks carried out for our main result. By making $x_A\!=\!\bar{x}_A$, it follows that ${\bf r}_2\!=\!{\bf r}_{\bar{2}}$, which leads to the first three terms of \eqref{largenc1b} vanishing (due to $\text{Tr}\{V_{\bar{2}}t^aV^{\dagger}_2\}\rightarrow\text{Tr}\{t^a\}=0$).
As we also have ${\bf r}_1\!=\!{\bf r}_{\bar{1}}$, the only surviving term reduces to:
\begin{align}
\text{\eqref{largenc1b}}\!\underset{x_A=\bar{x}_A}{\longrightarrow}\frac{{\cal C}_f}{N_c}\int_{{\bf z}_1}\!\int {\cal D} {\bf r}_3 \,{\cal D} {\bf r}_{\bar{3}}\left\langle \left(S_{13}\right)^2\right\rangle_{(\bar{x}_g,x_g)}\!\!\left\langle \left(S_{3\bar{3}}\right)^2\right\rangle_{(L,\bar{x}_g)}.\label{eq:check1}
\end{align}
It is straightforward to check that \eqref{largenc1} yields the same result in this limit. However, as mentioned above, the absence of adjoint Wilson lines in these expressions makes it difficult to connect them to the BDMPS-Z spectrum. This comparison comes more naturally when we take the instantaneous antenna formation limit over the `leading-$N_c$ + some subleading corrections' result shown in \eqref{largenc}. In doing so, the first four terms mutually cancel, with the last one reducing to:
\begin{align}
\text{\eqref{largenc}}\!\underset{x_A=\bar{x}_A}{\longrightarrow}\frac{{\cal C}_a}{4N_c}\int_{{\bf z}_1}\left\langle\text{Tr}\left\{ {\cal G}U^{\dagger}_1\right\}\right\rangle_{(\bar{x}_g,x_g)}
\!\!\left\langle \text{Tr}\left\{ {\cal G}{\cal G}^{\dagger} \right\}\right\rangle_{(L,\bar{x}_g)}.\label{eq:check2}
\end{align}
Note that both Eqs.~\!(\ref{eq:check1}) and (\ref{eq:check2}) scale like $N_c/2$.

We can also examine what happens when we `push' the gluon emission outside the medium. In our simplified treatment, this is implemented by replacing the propagators in regions $(\bar{x}_g,x_g)$ and $(L,\bar{x}_g)$ with the identity. It is straightforward to verify that, by doing this, the first four terms of \eqref{largenc1} cancel and the last term reduces to:
\begin{align}
\text{\eqref{largenc1}}\rightarrow {\cal C}_f \left\langle S_{12}\right\rangle_{(\bar{x}_A,x_A)}\left\langle Q_{12\bar{2}\bar{1}} \right\rangle_{(L',\bar{x}_A)}
\underset{N_c\uparrow}{\approx} C_FN_c\left\langle S_{12}\right\rangle_{(\bar{x}_A,x_A)}\left\langle Q_{12\bar{2}\bar{1}} \right\rangle_{(L',\bar{x}_A)},
\end{align}
in agreement with the result obtained in \cite{Abreu:2024wka}.

\section{Two limiting cases}
\label{sec:limits}
In the previous sections, we have suggested that color correlations with the spectator leg of the antenna (the antiquark, in our case) are the main driver of the deviations from the BDMPS-Z spectrum featured in our results. To further support this conclusion, we now examine two special limits.

\begin{figure}
\centering
\includegraphics[width=0.483\textwidth]{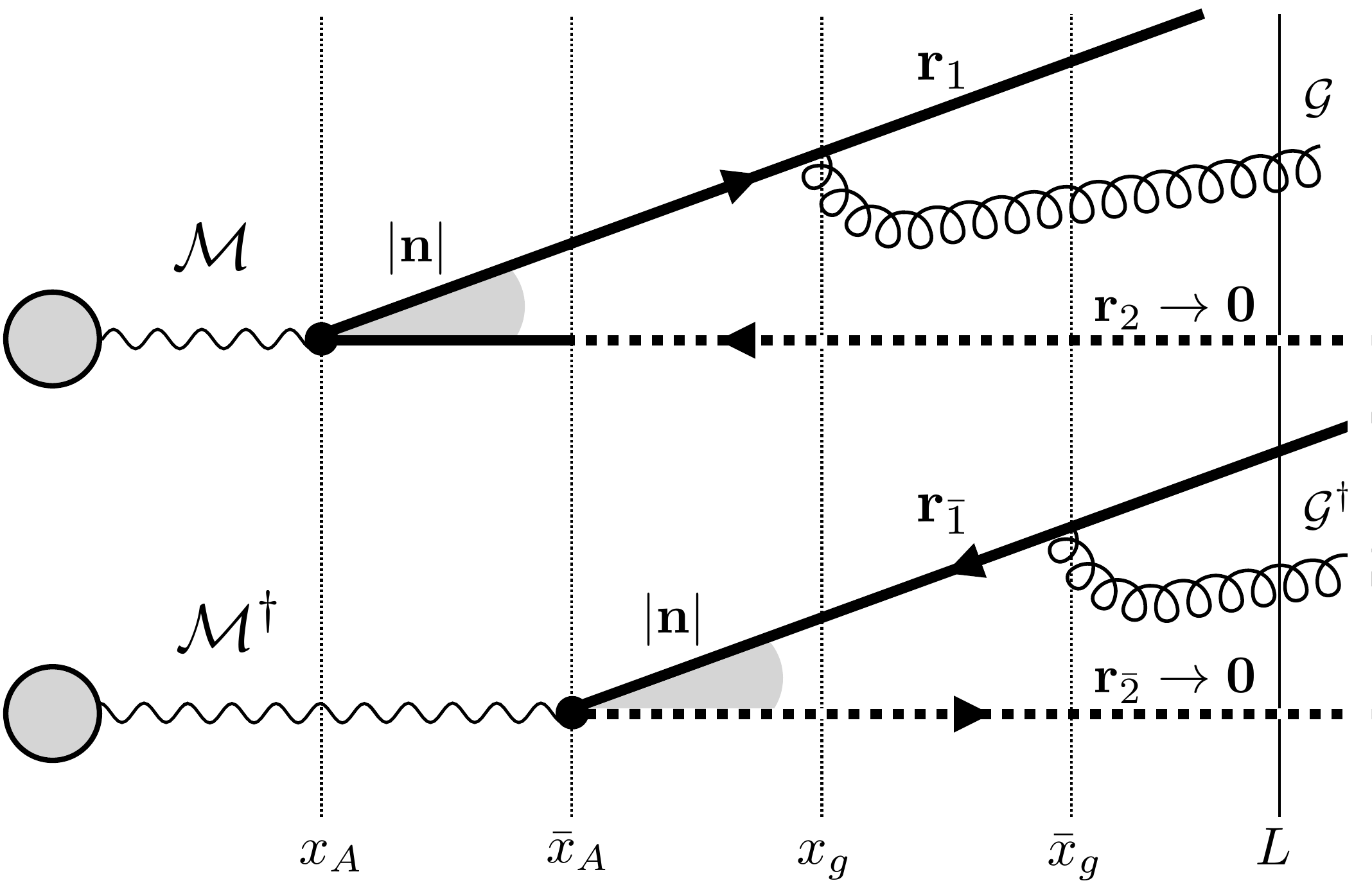}\hspace{0.4cm}
\includegraphics[width=0.483\textwidth]{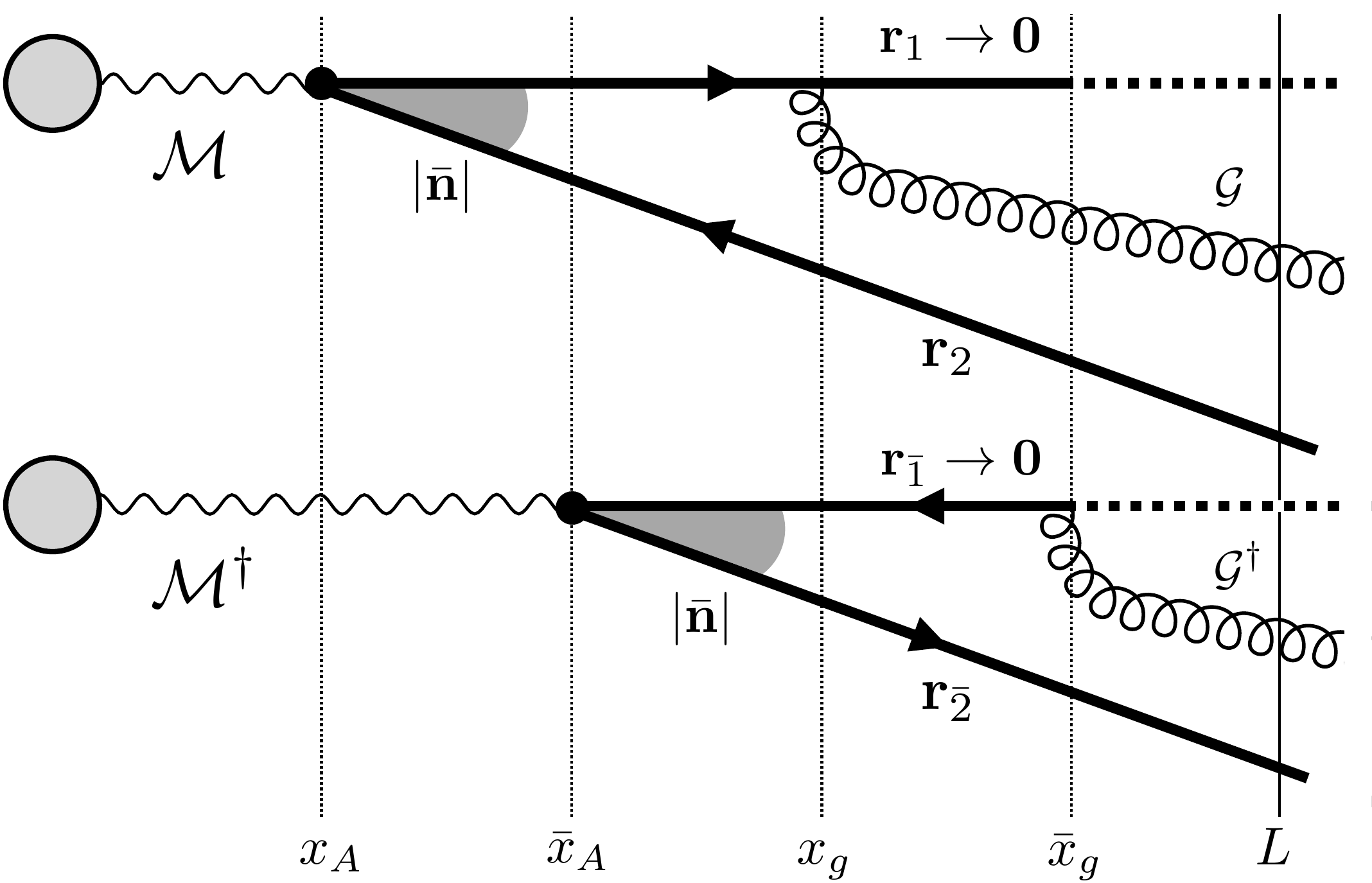}
\caption{Diagrams depicting the physical cases analyzed in \secref{sec:limits}. The left panel represents the case where the non-emitting prong of the antenna carries most of the energy of the photon, thus propagating along a completely flat (i.e.\ constant) transverse path. The right panel represents the opposite situation, where the spectator follows a tilted trajectory and the emitter remains at ${\bf 0}$ throughout the evolution. The dashed thick lines represent Wilson lines that mutually cancel upon multiplication of the amplitudes.}
\label{fig:limit1}
\end{figure}

\subsection{Hard spectator}
\label{sec:limit1}
We start with the limit where the energy-sharing fraction of the antiquark is approximately 1, meaning that it carries almost the totality of the energy of the parent photon\footnote{Despite the emitter leg of the antenna carrying a negligible fraction of the photon energy, this energy may still be assumed to be arbitrarily large. This allows us to describe the emitter by means of an eikonal propagator and treat the emitted gluon as soft even in this extreme case.}. This scenario is depicted on the left panel of \figref{fig:limit1}. In this regime, the `tilt' that characterizes the antiquark's eikonal trajectories in amplitude and conjugate amplitude is negligible ($\bar{\bf n}\rightarrow0$) and can thus be treated as a straight, flat line: ${\bf r}_2\approx{\bf r}_{\bar{2}}\rightarrow{\bf 0}$. As a result, the corresponding Wilson lines cancel almost entirely, surviving only within the $(\bar{x}_A,x_A)$ region. The remaining evolution consists of a BDMPS-Z-like diagram characterized by a transverse `gap' between amplitude and conjugate amplitude\footnote{An equivalent diagram was previously analyzed in \cite{Wu:2011kc}.}. The corresponding medium average reduces to:
\begin{align}
\left\langle...\right\rangle\!\underset{\bar{\bf n}\rightarrow0}{\longrightarrow}\frac{2}{(N_c^2-1)}\int_{{\bf z}_1}\hspace{11.5cm}\nonumber\\
\left\langle S_{10}\right\rangle_{(\bar{x}_A,x_A)} \left\langle S_{1\bar{1}}\right\rangle_{(x_g,\bar{x}_A)} \left\langle {\cal G}^{ab} \text{Tr}\left\{V^{\dagger}_{\bar{1}}t^aV_1t^b\right\}\right\rangle_{(\bar{x}_g,x_g)} \left\langle {\cal G}^{cd}\text{Tr}\left\{t^dt^eV^{\dagger}_{\bar{1}}V_1\right\}{\cal G}^{\dagger ec}\right\rangle_{(L,\bar{x}_g)}.
\end{align}
In this expression, we identify a pair of dipole functions encoding the color exchanges happening during and after the antenna splitting ($\langle S_{10}\rangle_{(\bar{x}_A,x_A)}$ and $\langle S_{1\bar{1}}\rangle_{(x_g,\bar{x}_A)} $, respectively), followed by generalizations of the gluon emission kernel and the momentum broadening factor. The specific modifications consist of a $\text{Tr}\{V^{\dagger}_{\bar{1}}t^aV_1t^b \}$ factor in the $(\bar{x}_g,x_g)$-region correlator (which reduces to an adjoint Wilson line in the instantaneous antenna formation limit) and a $\text{Tr}\{t^dt^eV^{\dagger}_{\bar{1}}V_1 \}$ factor in the $(L,\bar{x}_g)$-region correlator (which vanishes in the same limit). Despite these features, it is readily seen that the result of the medium average has simplified dramatically with respect to the general case analyzed in the previous section. The reason for such a simplification is the near-complete cancelation of spectator particle propagators. To further illustrate this point, in the next section we focus on the opposite case.

\subsection{Hard emitter}
\label{sec:limit2}
We now examine the situation where the gluon-emitting particle (the quark, in our case) carries most of the energy of the photon. This limit is equivalent to taking ${\bf n}\rightarrow 0$, which in turn leads to ${\bf r}_1\approx{\bf r}_{\bar{1}}\rightarrow{\bf 0}$. When we do this, \eqref{finalrad} becomes:
\begin{align}
\langle...\rangle\underset{{\bf n}\rightarrow0}{\longrightarrow}{\cal C}\int_{{\bf z}_1} \left\langle\text{Tr}\left\{V_0V^{\dagger}_2\right\}\right\rangle_{\scriptscriptstyle (\bar{x}_A,x_A)} \left\langle {\bold A}_0\right\rangle_{\scriptscriptstyle (x_g,\bar{x}_A)} \!\cdot{\bold g}_2\cdot\left\langle {\cal G}^{ab} {\bold B}^{ba}_0\right\rangle_{\scriptscriptstyle (\bar{x}_g,x_g)} \!\!\cdot{\bold g}_4\cdot\left\langle {\cal G}^{cd}{\bold C}^{de}_0{\cal G}^{\dagger ec}\right\rangle_{\scriptscriptstyle (L,\bar{x}_g)},
\end{align}\vspace{-0.5em}
with:
\begin{align}
{\bold A}^{\text{T}}\underset{{\bf n}\rightarrow0}{\longrightarrow}&\, \begin{pmatrix}  \text{Tr}\{ V_0 V^{\dagger}_2\}\text{Tr}\{ V_{\bar{2}} V^{\dagger}_{\bar{0}}\} \\   \text{Tr}\{V^{\dagger}_2V_{\bar{2}} \}\end{pmatrix}\equiv {\bold A}^{\text{T}}_0\\
\left({\bold B}^{ba}\right)^{\text{T}} \underset{{\bf n}\rightarrow0}{\longrightarrow} &\, \begin{pmatrix} \text{Tr}\{ t^b V^{\dagger}_2V_{\bar{2}}t^{b'}\}U^{\dagger b'a}_0 & U^{\dagger ba}_0\text{Tr}\{ V_{\bar{2}} V^{\dagger}_{2} \}/2 \\   \text{Tr}\{ V_0 t^bV^{\dagger}_2 \}\text{Tr}\{V^{\dagger}_{0}t^aV_{\bar{2}} \} &  U^{\dagger bb'}_0\text{Tr}\{ t^{b'}t^a V_{\bar{2}} V^{\dagger}_{2}\}\\  \text{Tr}\{ t^bV^{\dagger}_2 t^a V_0 \}\text{Tr}\{ V_{\bar{2}} V^{\dagger}_{0} \} & U^{\dagger bb'}_0\text{Tr}\{ t^{b'}V_{\bar{2}} V^{\dagger}_{2}t^a\}\\  \text{Tr}\{ t^bV^{\dagger}_{2}t^a V_{\bar{2}}  \} &   0 \end{pmatrix}\equiv \left({\bold B}^{ba}_0\right)^{\text{T}}\\
{\bold C}^{de}  \underset{{\bf n}\rightarrow0}{\longrightarrow} &\, \begin{pmatrix}\delta^{de}\text{Tr}\{ V^{\dagger}_{2}V_{\bar{2}}  \}/2  \\  \text{Tr}\{  t^dt^e V^{\dagger}_{2} V_{\bar{2}}  \}  \\  \text{Tr}\{ t^d V^{\dagger}_{2}V_{\bar{2}}t^e   \}  \\ 0 \end{pmatrix}\equiv  {\bold C}^{de}_0.
\end{align}
Remarkably, while the limit analyzed in \secref{sec:limit1} leads to a simple result straightforwardly (even without resorting to the large-$N_c$ limit), the present case retains much of the complexity of the general problem. This remains true even when we apply ${\bf n}\rightarrow 0$ over \eqref{largenc}, which becomes:
\begin{align}\label{largenc2}
\text{\eqref{largenc}}\!\underset{{\bf n}\rightarrow0}{\longrightarrow}{\cal C}_a\!\int_{{\bf z}_1}\hspace{11.8cm}\nonumber\\
\left\langle S_{02}\right\rangle_{\scriptscriptstyle (\bar{x}_A,x_A)}\!\Bigg[\!\left\langle S_{02} S_{\bar{2}\bar{0}}\right\rangle_{\scriptscriptstyle (x_g,\bar{x}_A)}\bigg(\left\langle{\cal G}^{ab}\text{Tr}\{ t^bV^{\dagger}_2 t^a V_0 \}\,S_{\bar{2}\bar{0}}\right\rangle_{\!\! \scriptscriptstyle (\bar{x}_g,x_g)}
\!\!\left\langle{\cal G}^{cd}\text{Tr}\{  t^d V^{\dagger}_{2}V_{\bar{2}} t^e  \}{\cal G}^{\dagger ec}\right\rangle_{\!\! \scriptscriptstyle (L,\bar{x}_g)}\nonumber\\
-\frac{1}{2}\left\langle{\cal G}^{ab}\text{Tr}\{ t^bV^{\dagger}_2 t^a V_0 \}\,S_{\bar{2}\bar{0}}\right\rangle_{\!\! \scriptscriptstyle (\bar{x}_g,x_g)}
\!\!\left\langle{\cal G}^{cd}S_{\bar{2}2}\,{\cal G}^{\dagger dc}\right\rangle_{\!\! \scriptscriptstyle (L,\bar{x}_g)}\nonumber\\
+\frac{1}{2}\left\langle{\cal G}^{ab}\text{Tr}\{ t^b V^{\dagger}_2V_{\bar{2}} t^{b'}\}U^{\dagger b'a}_0\right\rangle_{\!\! \scriptscriptstyle (\bar{x}_g,x_g)}
\!\!\left\langle{\cal G}^{cd}S_{\bar{2}2}\,{\cal G}^{\dagger dc}\right\rangle_{\!\!\scriptscriptstyle (L,\bar{x}_g)}\nonumber\\
-\frac{1}{4}\left\langle{\cal G}^{ab}U^{\dagger ba}_0\,S_{\bar{2}2}\right\rangle_{\!\! \scriptscriptstyle (\bar{x}_g,x_g)}
\!\!\left\langle{\cal G}^{cd}S_{\bar{2}2}\,{\cal G}^{\dagger dc}\right\rangle_{\!\! \scriptscriptstyle (L,\bar{x}_g)}\bigg)\nonumber\\
+\left\langle S_{\bar{2}2}\right\rangle_{\scriptscriptstyle (x_g,\bar{x}_A)}\bigg(\frac{1}{4}\left\langle{\cal G}^{ab}U^{\dagger ba}_0\,S_{\bar{2}2}\right\rangle_{\!\! \scriptscriptstyle (\bar{x}_g,x_g)}
\!\!\left\langle{\cal G}^{cd}S_{\bar{2}2}\,{\cal G}^{\dagger dc}\right\rangle_{\!\! \scriptscriptstyle (L,\bar{x}_g)}\bigg)\Bigg].
\end{align}
Again, almost no simplification ensues from this limit.
This suggests that the spectator particle, through its nontrivial correlations with both the emitter and the gluon, is primarily responsible for significantly complicating the calculation.
This idea is further supported by the fact that one effectively recovers the BDMPS-Z spectrum in the limit where the spectator is distant from the emitter, previously discussed in \secref{sec:largenc}. In said limit, the first four terms of \eqref{largenc2} become exponentially suppressed with respect to the last one (see appendix~\ref{app:limit}), which at large $N_c$ reduces to:
\begin{align}
\text{\eqref{largenc2}}\rightarrow\frac{{\cal C}_a}{4}\int_{{\bf z}_1}\left\langle S_{02}\right\rangle_{\!(\bar{x}_A,x_A)}\!
\left\langle S_{\bar{2}2}\right\rangle_{\!(L,\bar{x}_A)}\!
\left\langle\text{Tr}\left\{{\cal G}U^{\dagger}_0\right\}\right\rangle_{\!\!(\bar{x}_g,x_g)}\!
\left\langle\text{Tr}\left\{{\cal G}{\cal G}^{\dagger}\right\}\right\rangle_{\!\!(L,\bar{x}_g)},
\end{align}
scaling like $N_c^2/2\approx C_FN_c$.
Essentially, this expression consists of a gluon emission kernel and a momentum broadening factor corrected by the survival probabilities of the color states of the antenna, $\langle S_{02} \rangle$; and the antiquark, $\langle S_{\bar{2}2} \rangle$.

\section{Conclusions}
\label{sec:conclusions}

We have revisited the calculation of the soft gluon emission spectrum off a singlet quark-antiquark antenna in the presence of a QCD medium, first undertaken in \cite{Mehtar-Tani:2011vlz,Casalderrey-Solana:2011ule,Mehtar-Tani:2012mfa}. In our approach, we removed one of the standard approximations adopted in the literature, where the antenna is assumed to be generated instantaneously. Previous steps in this direction were taken in \cite{Dominguez:2019ges,Isaksen:2023nlr}, which focused on medium-induced modifications to the antenna production; and \cite{Abreu:2024wka}, which additionally considered the antenna to radiate a soft gluon outside of the medium. In this work we have taken the emission to happen within the medium, thus incorporating medium effects during the formation of both antenna and gluon. We did this by focusing on the direct contribution to the corresponding squared matrix element.

Working in the limit where the antenna is formed instantaneously amounts to taking the $\gamma\rightarrow q\bar{q}$ splitting to happen at the same light-cone time in both amplitude and conjugate amplitude. With this setup, the calculation of the direct contribution to the squared amplitude effectively reduces to that of the well-known BDMPS-Z spectrum, due to the propagators representing the spectator (i.e.\ non-emitting) prong of the antenna canceling out. By contrast, in this paper we treated the antenna formation time as an additional variable of the calculation. This modification spoils the aforementioned cancelations, thus enabling the spectator parton to participate in the overall color exchange with the medium. This gives rise to new correlations that violate the `naive' picture of color decoherence obtained under the usual approximations.
In the specific case examined here, the spectator particle is the antiquark, but the same reasoning applies if the roles of quark and antiquark are exchanged.

Our result for the direct contribution to the $\gamma\rightarrow q\bar{q}g$ squared matrix element is presented compactly in \eqref{finalrad} using matrix notation. This expression differs from the BDMPS-Z spectrum both in the number of independent terms and in their structure. While the in-in contribution to the BDMPS-Z spectrum can be expressed as a single term involving two correlators, our result (written in fundamental representation) comprises 160 distinct terms, each containing four correlators.
In the large-$N_c$ limit, this proliferation of terms can be reduced to only four, which we collect in \eqref{largenc1b}. This more tractable expression allows for a clearer physical interpretation of each contribution. For instance, one of the terms corresponds to a configuration in which, when taking the large-$N_c$ limit, the spectator rotates its color independently from both emitter and gluon. This contribution thus describes a situation in which (except during the formation of the antenna itself, where quark and antiquark necessarily evolve in a dipole state), the spectator does not `see' the rest of the system. The resulting expression is a generalization of the BDMPS-Z spectrum, up to correction factors which encode the survival probability of the antenna and the spectator.

Conversely, the other three leading-$N_c$ terms correspond to configurations in which the spectator does correlate with the rest of the antenna. These correlations are restricted to limited light-cone time intervals, after which the system transitions to a decorrelated configuration via an effective gluon exchange. As a consequence of this exchange, as well as the prior correlations, these contributions become exponentially suppressed in the limit where the spectator is far from the rest of the system. Such a regime is realized for antennas with both a wide opening angle and a late gluon emission time. This argument suggests that the influence of the spectators should become increasingly important along the sequence of successive splittings in a jet, as these generally occur at smaller angles.

Finally, in \secref{sec:limits} we examine our results in the limit of extremely asymmetric photon splittings. We first consider the case in which the spectator leg of the antenna carries most of the energy. In this regime, its kinematics become trivial, meaning that it remains fixed at transverse position ${\bf 0}$ throughout its propagation across the medium. As a consequence, the corresponding Wilson lines cancel almost completely between the amplitude and the conjugate amplitude. Only during the formation time of the antenna does the spectator participate in the color exchange, giving rise to a decoherence phase. The subsequent evolution can then be interpreted as a generalization of the BDMPS-Z spectrum, in which the emitter propagates with a light-cone time offset between amplitude and conjugate amplitude. This `lag' induces an additional decoherence phase, together with generalizations of the gluon emission kernel and the momentum broadening factor. More challenging is the opposite limit, where the emitter particle carries most of the energy of the parent photon. We observe that, although some simplifications emerge in this regime, the overall unwieldiness of the color algebra involved in the general case largely persists. This is mainly due to the presence of spectator particles in the medium averaging procedure, which sustains the proliferation of non-trivial color correlations.

We have restricted ourselves to considering only the direct contribution to the squared amplitude of the $\gamma\rightarrow q\bar{q}g$ process. Our approach is purely analytical, leaving a numerical study for future work. We also plan to extend the scope of this analysis to other contributions to the squared matrix element, including interference terms as well as alternative orderings of the integration variables. Of particular interest is the case where the formation times of antenna and gluon overlap (previously studied in \cite{Arnold:2015qya,Casalderrey-Solana:2015bww} for different physical setups). Taken together, these contributions will enable the computation of the full emission spectrum, which will in turn allow us to assess the potential impact of our results on the formulation of in-medium parton cascades. From an analytical perspective, it will also be particularly interesting to compare with earlier studies such as \cite{Dominguez:2019ges,Abreu:2024wka}, in which medium effects are naturally incorporated into compact expressions as corrections to the vacuum spectrum.

\acknowledgments
This work has been supported by European Research Council under project ERC-2018-ADG-835105 YoctoLHC; by Maria de Maeztu excellence unit grant CEX2023-001318-M and project PID2023-152762NB-I00 funded by MICIU/AEI/10.13039/501100011033; by ERDF/EU; and by Xunta de Galicia (CIGUS Network of Research Centres). This project also received funding from the Ecole Polytechnique Foundation.
\appendix

\section{Medium averages}
\label{app:medav}
In this appendix, we provide details of the calculation leading to \eqref{finalrad}. Our starting point is the color correlator featured in the squared matrix element \eqref{me1}, which we rewrite here using a simplified notation:
\begin{align}\label{corr1}
\hspace{-0.7em}\Big\langle\big[{\cal G^{\dagger}}(\bar{x}_g,L){\cal G}(L,x_g)\big]^{b'b}
\text{Tr}\big\{V^{\dagger}_{\bar{1}}(\bar{x}_A,\bar{x}_g)t^{b'}V^{\dagger}_{\bar{1}}(\bar{x}_g,L)V_1(L,x_g)t^bV_1(x_g,x_A)V^{\dagger}_2(x_A,L)V_{\bar{2}}(L,\bar{x}_A)\big\}\Big\rangle.
\end{align}
We consider the contribution where $x_A<\bar{x}_A<x_g<\bar{x}_g<L$. As we assume the medium averages to be local, \eqref{corr1} can be factorized into distinct light-cone time regions. To do this, we split the Wilson lines and gluon propagators appropriately:
\begin{align}\label{eq:trace}
\int_{{\bf z}_1}\Big\langle \, &{\cal G}^{\dagger b'a}(\bar{x}_g,L){\cal G}^{ac}(L,\bar{x}_g){\cal G}^{cb}(\bar{x}_g,x_g)\text{Tr}\Big\{V^{\dagger}_{\bar{1}}(\bar{x}_A,x_g)V^{\dagger}_{\bar{1}}(x_g,\bar{x}_g)t^{b'}V^{\dagger}_{\bar{1}}(\bar{x}_g,L)\nonumber\\
& V_1(L,\bar{x}_g)V_1(\bar{x}_g,x_g)t^bV_1(x_g,\bar{x}_A)V_1(\bar{x}_A,x_A)V^{\dagger}_2(x_A,\bar{x}_A)V^{\dagger}_2(\bar{x}_A,x_g)V^{\dagger}_2(x_g,\bar{x}_g)\nonumber\\
& V^{\dagger}_2(\bar{x}_g,L)V_{\bar{2}}(L,\bar{x}_g)V_{\bar{2}}(\bar{x}_g,x_g)V_{\bar{2}}(x_g,\bar{x}_A)\Big\}\Big\rangle.
\end{align}
Note that, to split a gluon propagator, one must apply the composition law \eqref{Gsplit}, which introduces an integration over the intermediate position ${\bf z}_1$. Further, since we are presently concerned only with the color algebra involved in the correlator, we explicitly separate the path integrals of the gluon propagators from their Wilson line parts:
\begin{align}\label{factorize}
{\cal G}^{ab}(\bar{t},{\bf y};t,{\bf x}|E)=&\,\int^{{\bf r}(\bar{t})={\bf y}}_{{\bf r}(t)={\bf x}} {\cal D} {\bf r}(s) \, \exp\left\{ \frac{iE}{2}\int \diff s \left( \frac{\diff{\bf r}}{\diff s}\right)^2\right\} U^{ab}(\bar{t},t;\left[{\bf r(s)}\right]),
\end{align}
which we then transform into the fundamental representation by applying the following identity:
\begin{align}\label{eq:ident}
U^{ab}t^b_{ij}&\,=\left[V^{\dagger}t^aV\right]_{ij}.
\end{align}
This results in:
\begin{align}
U_{\bar{3}}^{\dagger b'a}(\bar{x}_g,L)t^{b'}_{ij}&=U_{\bar{3}}^{ab'}t^{b'}_{ij}=\left[ V^{\dagger}_{\bar{3}}(\bar{x}_g,L)t^{a}V_{\bar{3}}(L,\bar{x}_g)\right]_{ij}\\
U_3^{ab}(L,x_g)t^b_{kl}&=\left[ V^{\dagger}_{3}t^{a}V_{3}\right]_{kl}=\left[ V^{\dagger}_{3}(x_g,\bar{x}_g)V^{\dagger}_{3}(\bar{x}_g,L)t^{a}V_{3}(L,\bar{x}_g)V_{3}(\bar{x}_g,x_g)\right]_{kl},
\end{align}
where the Wilson line paths ${\bf r}_{3}$ and ${\bf r}_{\bar{3}}$ are path-integrated over. By substituting these expressions into \eqref{eq:trace}, applying the Fierz identity (\eqref{eq:fierz}), and rearranging into light-cone time regions, we eventually arrive at:
\begin{align}\label{eq:start}
\frac{1}{2N_c}\Big\langle &\!\left(\text{Tr}\left\{V_1V^{\dagger}_2\right\}\right)_{\!\scriptscriptstyle (\bar{x}_A,x_A)}\!\!\left(\left[ V_1 V^{\dagger}_2 \right]_{ij} \left[ V_{\bar{2}} V^{\dagger}_{\bar{1}}\right]_{kl} \right)_{\!\scriptscriptstyle (x_g,\bar{x}_A)}\!\!\!    \left( \left[ V_1 V^{\dagger}_3\right]_{mn}\!(V^{\dagger}_2)_{jI}(V_{\bar{2}})_{Jk}(V^{\dagger}_{\bar{1}})_{lI_1}(V_3)_{I_2i}\right)_{\!\scriptscriptstyle (\bar{x}_g,x_g)}\nonumber\\
&\,\left( \left[ V^{\dagger}_3V_{\bar{3}}V^{\dagger}_{\bar{1}}V_1 \right]_{nm} \left[ V^{\dagger}_2V_{\bar{2}}\right]_{IJ} \left[ V^{\dagger}_{\bar{3}}V_{3}\right]_{I_1I_2}\right)_{(L,\bar{x}_g)}\Big\rangle\nonumber\\
-\frac{1}{2N^2_c}\Big\langle &\!\left(\text{Tr}\left\{V_1V^{\dagger}_2\right\}\right)_{\!\scriptscriptstyle (\bar{x}_A,x_A)}\!\!\left(\left[ V_1 V^{\dagger}_2 \right]_{ij} \left[ V_{\bar{2}} V^{\dagger}_{\bar{1}}\right]_{kl} \right)_{\!\scriptscriptstyle(x_g,\bar{x}_A)}\!\!\! \left( (V^{\dagger}_2)_{jI}(V_{\bar{2}})_{Jk}(V^{\dagger}_{\bar{1}})_{ln}(V_1)_{mi}\right)_{\!\scriptscriptstyle(\bar{x}_g,x_g)}\nonumber\\
&\, \left( \left[ V^{\dagger}_{\bar{1}}V_1\right]_{nm} \left[V^{\dagger}_2V_{\bar{2}}\right]_{IJ}\right)_{\!\scriptscriptstyle(L,\bar{x}_g)}\Big\rangle.
\end{align}
Note that we are considering quark, antiquark and gluon to be in definite color states (respectively labeled as $i$, $m$, and $a$ in \eqref{amp}) that we sum over as we square the amplitude. \eqref{eq:start} is thus in a color-singlet state. Further, we assume that no color is exchanged with the medium on average (see \eqref{2pf}), which implies that the system remains a color singlet throughout the interaction. It then follows that, for each light-cone time region, all color indices must be contracted into traces; in other words, each color vector must be projected onto a linear combination of singlets. This has already been done in the expression above for the trivial case of the $(\bar{x}_A,x_A)$ region, where we have a two-point function and therefore the only choice is:
\begin{align}
\left\langle\left[V_1V^{\dagger}_2\right]_{ij}\right\rangle=\frac{1}{N_c}\delta^{ij}\left\langle\text{Tr}\left\{V_1V^{\dagger}_2\right\}\right\rangle.
\end{align}
This is, however, not so simple for the remaining regions. Let us first focus on the $(x_g,\bar{x}_A)$ region, which contains four open indices.
In this case, we need to project onto the following orthonormal basis \cite{Fukushima:2007dy,Fukushima:2017mko,Barata:2024bqp}:
\begin{align}
s^{ijkl}_1&\!=\frac{1}{N_c}\delta^{ij}\delta^{kl}\label{eq:basis1}\\
s^{ijkl}_2&\!=\frac{1}{\sqrt{N^2_c-1}}\left(\delta^{il}\delta^{jk}-\frac{1}{N_c}\delta^{ij}\delta^{kl}\right).\label{eq:basis2}
\end{align}
These color vectors can be used to project the correlators of region $(x_g,\bar{x}_A)$ onto products of traces, resulting in \cite{Barata:2024bqp}:
\begin{align}
\left\langle\left[ V_1 V^{\dagger}_2 \right]_{ij}\right. &\!\left. \left[ V_{\bar{2}} V^{\dagger}_{\bar{1}}\right]_{kl} \right\rangle_{(x_g,\bar{x}_A)}=\frac{1}{N_c}\left\langle \text{Tr}\left\{V_1 V^{\dagger}_2\right\}\text{Tr}\left\{V_{\bar{2}} V^{\dagger}_{\bar{1}}\right\} \right\rangle s_1^{ijkl}\nonumber\\
&+\frac{1}{\sqrt{N_c^2-1}}\left(\left\langle  \text{Tr}\left\{V_1 V^{\dagger}_2V_{\bar{2}} V^{\dagger}_{\bar{1}}\right\} \right\rangle -\frac{1}{N_c}\left\langle  \text{Tr}\left\{V_1 V^{\dagger}_2\right\}\text{Tr}\left\{V_{\bar{2}} V^{\dagger}_{\bar{1}}\right\} \right\rangle\right)s_2^{ijkl}.
\end{align}
The same can be done for region $(L,\bar{x}_g)$ in the $N_c$-suppressed term of \eqref{eq:start}. Overall, the two-dimensional basis suffices to fully project this term onto singlets. Unfortunately, it is not enough for the other contribution, where we encounter regions that contain six open indices. An orthonormal basis for the corresponding projectors was obtained in \cite{Lappi:2020srm}:
\begingroup
\allowdisplaybreaks
\begin{align}
s^{ijklmn}_1&\!=\frac{1}{\sqrt{N_c^3}}\delta^{ij}\delta^{kl}\delta^{mn}\\
s^{ijklmn}_2&\!=\frac{1}{\sqrt{N_c(N^2_c-1)}}\left(\delta^{il}\delta^{jk}-\frac{1}{N_c}\delta^{ij}\delta^{kl}\right)\delta^{mn}\\
s^{ijklmn}_3&\!=\frac{1}{\sqrt{N_c(N^2_c-1)}}\left(\delta^{in}\delta^{jm}-\frac{1}{N_c}\delta^{ij}\delta^{mn}\right)\delta^{kl}\\
s^{ijklmn}_4&\!=\frac{1}{\sqrt{N_c(N^2_c-1)}}\left(\delta^{kn}\delta^{lm}-\frac{1}{N_c}\delta^{mn}\delta^{kl}\right)\delta^{ij}\\
s^{ijklmn}_5&\!=\frac{1}{\sqrt{2N_c(N^2_c-1)}}\left(\delta^{il}\delta^{kn}\delta^{mj}-\delta^{in}\delta^{ml}\delta^{jk}\right)
\end{align}
\endgroup
\begin{align}
s^{ijklmn}_6&\!=\sqrt{\frac{N_c}{2(N_c^2-4)(N_c^2-1)}}\Bigg(\delta^{il}\delta^{kn}\delta^{mj}+\delta^{in}\delta^{ml}\delta^{jk}\nonumber\\
&\hspace{8em}-\frac{2}{N_c}\left(\delta^{il}\delta^{jk}\delta^{mn}+\delta^{in}\delta^{jm}\delta^{kl}+\delta^{kn}\delta^{lm}\delta^{ij}\right)+\frac{4}{N_c^2}\delta^{ij}\delta^{kl}\delta^{mn}\Bigg).
\end{align}
Projecting onto this basis and multiplying the contributions from each region yields a total of 160 terms. However, these can be expressed compactly in matrix notation as:
\begin{align}\label{fundrep}
\frac{1}{2N^3_c(N_c^2-1)^2(N_c^2-4)} \int_{{\bf z}_1}\left\langle\text{Tr}\left\{V_1V^{\dagger}_2\right\}\right\rangle_{\!\scriptscriptstyle (\bar{x}_A,x_A)} \left\langle {\bold A}\right\rangle_{\!\scriptscriptstyle (x_g,\bar{x}_A)}\cdot{\bold g}_2\cdot  \left\langle {\bold B}_f\right\rangle_{\!\scriptscriptstyle (\bar{x}_g,x_g)}\cdot {\bold G}_6\cdot \left\langle{\bold C}_f\right\rangle_{\!\scriptscriptstyle (L,\bar{x}_g)},
\end{align}
where ${\bold A}$ is defined as in the main body of the manuscript, and ${\bold B}_f$, ${\bold C}_f$ are given by:
\begin{align}
{\bold B}^{\text{T}}_f \!=\! \begin{pmatrix}  \text{Tr}\{ V_1 V^{\dagger}_3\}\text{Tr}\{ V_3 V^{\dagger}_2V_{\bar{2}} V^{\dagger}_{\bar{1}}\} \!\!&\!\!  \text{Tr}\{ V_1 V^{\dagger}_3 \}\text{Tr}\{ V_3 V^{\dagger}_{\bar{1}} \}\text{Tr}\{ V_{\bar{2}} V^{\dagger}_{2} \} \\   \text{Tr}\{ V_3 V^{\dagger}_2 V_1 V^{\dagger}_3 V_{\bar{2}} V^{\dagger}_{\bar{1}}\} \!\!&\!\!  \text{Tr}\{ V_3 V^{\dagger}_{\bar{1}} \}\text{Tr}\{ V_1 V^{\dagger}_3 V_{\bar{2}} V^{\dagger}_{2}\}\\   \text{Tr}\{ V^{\dagger}_2 V_{\bar{2}}V^{\dagger}_{\bar{1}} V_1\} \!\!&\!\!  \text{Tr}\{ V_1 V^{\dagger}_{\bar{1}} \}\text{Tr}\{ V_{\bar{2}} V^{\dagger}_{2}\}\\   \text{Tr}\{ V_1 V^{\dagger}_3 \}\text{Tr}\{ V_3 V^{\dagger}_2 \}\text{Tr}\{ V_{\bar{2}} V^{\dagger}_{\bar{1}} \} \!\!&\!\!  \text{Tr}\{ V_1 V^{\dagger}_3\}\text{Tr}\{ V_3 V^{\dagger}_{\bar{1}}V_{\bar{2}} V^{\dagger}_{2}\}\\   \text{Tr}\{ V_3 V^{\dagger}_2 \}\text{Tr}\{ V_1 V^{\dagger}_3 V_{\bar{2}} V^{\dagger}_{\bar{1}}\} \!\!&\!\!   \text{Tr}\{ V_3 V^{\dagger}_{\bar{1}} V_1 V^{\dagger}_3 V_{\bar{2}} V^{\dagger}_{2}\}\\   \text{Tr}\{ V_1 V^{\dagger}_2 \}\text{Tr}\{ V_{\bar{2}} V^{\dagger}_{\bar{1}}\} \!\!&\!\!  \text{Tr}\{ V^{\dagger}_{\bar{1}} V_{\bar{2}}V^{\dagger}_{2} V_1\} \end{pmatrix}\label{Bfund}\\
{\bold C}_f \!=\! \begin{pmatrix}  \text{Tr}\{ V^{\dagger}_3V_{\bar{3}} V^{\dagger}_{\bar{1}} V_1\}\text{Tr}\{ V^{\dagger}_2 V_{\bar{2}} \}\text{Tr}\{ V^{\dagger}_{\bar{3}} V_{3}\} \\    \text{Tr}\{ V^{\dagger}_3V_{\bar{3}} V^{\dagger}_{\bar{1}} V_1V^{\dagger}_2 V_{\bar{2}} \}\text{Tr}\{ V^{\dagger}_{\bar{3}} V_{3}\}  \\   \text{Tr}\{ V^{\dagger}_2 V_{\bar{2}} \}\text{Tr}\{ V^{\dagger}_{\bar{1}} V_{1}\} \\   \text{Tr}\{ V^{\dagger}_3V_{\bar{3}} V^{\dagger}_{\bar{1}} V_1\}\text{Tr}\{ V^{\dagger}_2V_{\bar{2}} V^{\dagger}_{\bar{3}} V_3\} \\    \text{Tr}\{ V^{\dagger}_3V_{\bar{3}} V^{\dagger}_{\bar{1}} V_1 V^{\dagger}_{\bar{3}} V_{3} V^{\dagger}_2 V_{\bar{2}} \}\\   \text{Tr}\{ V^{\dagger}_{\bar{1}}V_1 V^{\dagger}_2V_{\bar{2}}\}\end{pmatrix}.
\end{align}
The products between these vectors are mediated by ${\bold g}_2$ (defined in \eqref{metric}) and:
\begin{align}
{\bold G}_6=\left( \begin{array}{c|c}
   {\bold G}_{A} \,&\, {\bold G}_{B} \\
   \midrule
   {\bold G}_{B} \,&\, {\bold G}_{A} \\
\end{array}\right),
\end{align}
where:
\begin{align}
{\bold G}_{A} =\begin{pmatrix}
N_c^2-2 & -N_c & -N_c \\
-N_c & N_c^2-2  & 2 \\
-N_c & 2 & 2 
\end{pmatrix}\!,\,
{\bold G}_{B} =\begin{pmatrix}
-N_c & 2 & 2 \\
2 & -N_c  & -N_c \\
2 & -N_c & -4/N_c 
\end{pmatrix}\!.
\end{align}
In order to have a better grasp of the relation between this result and the BDMPS-Z spectrum, it is convenient to transform the Wilson lines along the gluon paths (labeled $3$ or $\bar{3}$, corresponding to the amplitude and conjugate amplitude, respectively) back to adjoint representation. We outline this procedure in appendix~\ref{app:adjoint}.

\section{Transforming (back) to adjoint representation}
\label{app:adjoint}
In this appendix, we recast the result derived in appendix~\ref{app:medav} in the form presented in the main text. Starting with \eqref{Bfund}, we perform the following transformation:
\begin{align}
{\bf B}_f={\cal B}\cdot\text{S}^{-1},
\end{align}
where:
\begin{align}
\hspace{-1em}\text{S}\!=\!\begin{pmatrix}
1/2 & 0 & 0 & 0 & 0 & 0 \\
0 & 1/2  & 0 & 0 & 0 & 0 \\
-\frac{1}{2N_c} & 0 & 1 & 0 & -\frac{1}{2N_c} & 0 \\
0 & 0 & 0 & 1/2 & 0 & 0 \\
0 & 0 & 0 & 0 & 1/2 & 0 \\
0 & -\frac{1}{2N_c} & 0 & -\frac{1}{2N_c} & 0 & 1 
\end{pmatrix}
\end{align}
and ${\cal B^{\text{T}}}=\left( \begin{array}{c|c}
   {\cal B}_A \,&\, {\cal B}_B
\end{array}\right)$, with:
\begin{align}\label{calB1}
{\cal B}_A\!=\!&\,\begin{pmatrix} \frac{1}{2}\left\langle\text{Tr}\{ V_1 V^{\dagger}_3\}\text{Tr}\{ V_3 V^{\dagger}_2V_{\bar{2}} V^{\dagger}_{\bar{1}}\}\right\rangle - \frac{1}{2N_c}\left\langle \text{Tr}\{ V^{\dagger}_2 V_{\bar{2}}V^{\dagger}_{\bar{1}} V_1\} \right\rangle  \\  \frac{1}{2}\left\langle \text{Tr}\{ V_3 V^{\dagger}_2 V_1 V^{\dagger}_3 V_{\bar{2}} V^{\dagger}_{\bar{1}}\}\right\rangle -\frac{1}{2N_c}\left\langle \text{Tr}\{ V_1 V^{\dagger}_2 \}\text{Tr}\{ V_{\bar{2}} V^{\dagger}_{\bar{1}}\}\right\rangle \\  \left\langle \text{Tr}\{ V^{\dagger}_2 V_{\bar{2}}V^{\dagger}_{\bar{1}} V_1\}\right\rangle \\  \frac{1}{2}\left\langle \text{Tr}\{ V_1 V^{\dagger}_3 \}\text{Tr}\{ V_3 V^{\dagger}_2 \}\text{Tr}\{ V_{\bar{2}} V^{\dagger}_{\bar{1}} \}\right\rangle -\frac{1}{2N_c}\left\langle \text{Tr}\{ V_1 V^{\dagger}_2 \}\text{Tr}\{ V_{\bar{2}} V^{\dagger}_{\bar{1}}\}\right\rangle\\  \frac{1}{2}\left\langle \text{Tr}\{ V_3 V^{\dagger}_2 \}\text{Tr}\{ V_1 V^{\dagger}_3 V_{\bar{2}} V^{\dagger}_{\bar{1}}\}\right\rangle- \frac{1}{2N_c}\left\langle \text{Tr}\{ V^{\dagger}_2 V_{\bar{2}}V^{\dagger}_{\bar{1}} V_1\} \right\rangle\\  \left\langle \text{Tr}\{ V_1 V^{\dagger}_2 \}\text{Tr}\{ V_{\bar{2}} V^{\dagger}_{\bar{1}}\}\right\rangle \end{pmatrix}
\end{align}
\begin{align}\label{calB2}
{\cal B}_B\!=\!&\,\begin{pmatrix}  \frac{1}{2}\left\langle \text{Tr}\{ V_1 V^{\dagger}_3 \}\text{Tr}\{ V_3 V^{\dagger}_{\bar{1}} \}\text{Tr}\{ V_{\bar{2}} V^{\dagger}_{2} \} \right\rangle -\frac{1}{2N_c}\left\langle \text{Tr}\{ V_1 V^{\dagger}_{\bar{1}} \}\text{Tr}\{ V_{\bar{2}} V^{\dagger}_{2}\}\right\rangle \\   \frac{1}{2}\left\langle \text{Tr}\{ V_3 V^{\dagger}_{\bar{1}} \}\text{Tr}\{ V_1 V^{\dagger}_3 V_{\bar{2}} V^{\dagger}_{2}\}\right\rangle - \frac{1}{2N_c}\left\langle \text{Tr}\{ V^{\dagger}_{\bar{1}} V_{\bar{2}}V^{\dagger}_{2} V_1\}\right\rangle\\ \left\langle \text{Tr}\{ V_1 V^{\dagger}_{\bar{1}} \}\text{Tr}\{ V_{\bar{2}} V^{\dagger}_{2}\}\right\rangle\\ \frac{1}{2}\left\langle \text{Tr}\{ V_1 V^{\dagger}_3\}\text{Tr}\{ V_3 V^{\dagger}_{\bar{1}}V_{\bar{2}} V^{\dagger}_{2}\}\right\rangle - \frac{1}{2N_c}\left\langle \text{Tr}\{ V^{\dagger}_{\bar{1}} V_{\bar{2}}V^{\dagger}_{2} V_1\}\right\rangle \\   \frac{1}{2}\left\langle \text{Tr}\{ V_3 V^{\dagger}_{\bar{1}} V_1 V^{\dagger}_3 V_{\bar{2}} V^{\dagger}_{2}\}\right\rangle -\frac{1}{2N_c}\left\langle \text{Tr}\{ V_1 V^{\dagger}_{\bar{1}} \}\text{Tr}\{ V_{\bar{2}} V^{\dagger}_{2}\}\right\rangle\\ \left\langle \text{Tr}\{ V^{\dagger}_{\bar{1}} V_{\bar{2}}V^{\dagger}_{2} V_1\}\right\rangle \end{pmatrix}.
\end{align}
The elements affected by this transformation can be rewritten by applying the Fierz identity to insert products of color matrices. Let us illustrate this procedure by considering the first element of \eqref{calB1}:
\begin{align}\label{invFI}
\frac{1}{2}\text{Tr}\{ V_1 V^{\dagger}_3\}\text{Tr}\{ V_3 V^{\dagger}_2V_{\bar{2}} V^{\dagger}_{\bar{1}}\} - \frac{1}{2N_c}\text{Tr}\{ V^{\dagger}_2 V_{\bar{2}}V^{\dagger}_{\bar{1}} V_1\}=\text{Tr}\{ V_1 V^{\dagger}_3 t^a V_3 V^{\dagger}_2V_{\bar{2}} V^{\dagger}_{\bar{1}} t^a\}.
\end{align}
Here we can use the identity $U^{ab}t^b_{ij}\!=\!\left[V^{\dagger}t^aV\right]_{ij}$ to obtain:
\begin{align}
\text{Tr}\{ V_1 V^{\dagger}_3 t^a V_3 V^{\dagger}_2V_{\bar{2}} V^{\dagger}_{\bar{1}} t^a\}=U^{ab}_3\text{Tr}\{ V_1 t^b V^{\dagger}_2V_{\bar{2}} V^{\dagger}_{\bar{1}} t^a\}.
\end{align}
By applying similar transformations to the remaining `Fierz-like' elements of Eqs.~(\ref{calB1}) and (\ref{calB2}), we eventually obtain:
\begin{align}\label{B_alt}
{\cal B}^{\text{T}} \!=\! \begin{pmatrix}  U^{ab}_3\text{Tr}\{ V_1 t^b V^{\dagger}_2V_{\bar{2}} V^{\dagger}_{\bar{1}}t^a \} \!&\!  U^{ab}_3\text{Tr}\{ t^bV^{\dagger}_{\bar{1}} t^a V_1 \}\text{Tr}\{ V_{\bar{2}} V^{\dagger}_{2} \} \\   U^{ab}_3 \text{Tr}\{ V_1 t^bV^{\dagger}_2 \}\text{Tr}\{V^{\dagger}_{\bar{1}}t^aV_{\bar{2}} \} \!&\!  U^{ab}_3 \text{Tr}\{ t^b V^{\dagger}_{\bar{1}} t^a V_{\bar{2}} V^{\dagger}_{2}V_1\}\\   \text{Tr}\{ V^{\dagger}_2 V_{\bar{2}}V^{\dagger}_{\bar{1}} V_1\} \!&\!  \text{Tr}\{ V_1 V^{\dagger}_{\bar{1}} \}\text{Tr}\{ V_{\bar{2}} V^{\dagger}_{2}\}\\   U^{ab}_3\text{Tr}\{ t^b V^{\dagger}_2 t^aV_1 \}\text{Tr}\{ V_{\bar{2}} V^{\dagger}_{\bar{1}} \} \!&\!  U^{ab}_3\text{Tr}\{ t^b V^{\dagger}_{\bar{1}}V_{\bar{2}} V^{\dagger}_{2}t^a V_1\}\\  U^{ab}_3 \text{Tr}\{ t^b V^{\dagger}_{2}t^a V_{\bar{2}} V^{\dagger}_{\bar{1}}V_1 \} \!&\!   U^{ab}_3 \text{Tr}\{ V_1 t^bV^{\dagger}_{\bar{1}} \}\text{Tr}\{V^{\dagger}_{2}t^aV_{\bar{2}} \}\\   \text{Tr}\{ V_1 V^{\dagger}_2 \}\text{Tr}\{ V_{\bar{2}} V^{\dagger}_{\bar{1}}\} \!&\!  \text{Tr}\{ V^{\dagger}_{\bar{1}} V_{\bar{2}}V^{\dagger}_{2} V_1\} \end{pmatrix}.
\end{align}
Similarly, by doing ${\bf C}_f=(\text{S}^{\text{T}})^{-1}\cdot{\cal C}$ and repeating the previous steps, we obtain:
\begin{align}\label{C_alt}
{\cal C} \!=\! \begin{pmatrix}  U^{cd}_3\text{Tr}\{  t^dt^e V^{\dagger}_{\bar{1}} V_1 \}\text{Tr}\{ V^{\dagger}_{2}V_{\bar{2}}  \}U^{\dagger ec}_{\bar{3}}  \\   U^{cd}_3\text{Tr}\{  t^dt^e V^{\dagger}_{\bar{1}} V_1 V^{\dagger}_{2} V_{\bar{2}}  \}U^{\dagger ec}_{\bar{3}}  \\   \text{Tr}\{ V^{\dagger}_2 V_{\bar{2}} \}\text{Tr}\{ V^{\dagger}_{\bar{1}} V_{1}\} \\   U^{cd}_3\text{Tr}\{   t^d V^{\dagger}_{2}V_{\bar{2}}t^e V^{\dagger}_{\bar{1}} V_1  \}U^{\dagger ec}_{\bar{3}}  \\  U^{cd}_3\text{Tr}\{ V_{\bar{2}}t^dV^{\dagger}_{2}  \}\text{Tr}\{  V_1t^e V^{\dagger}_{\bar{1}}\}U^{\dagger ec}_{\bar{3}} \\  \text{Tr}\{ V^{\dagger}_{\bar{1}}V_1 V^{\dagger}_2V_{\bar{2}}\}\end{pmatrix}\!.
\end{align}
The matrix mediating the products of these vectors then becomes sandwiched between the transformation matrices as $\text{S}^{-1}\!\!\cdot{\bf G}_6\cdot\!(\text{S}^{\text{T}})^{-1}$. The result reads:
\begin{align}
\text{S}^{-1}\!\!\cdot{\bf G}_6\cdot\!(\text{S}^{\text{T}})^{-1}\equiv\tilde{\bold G}_6=4\left( \begin{array}{c|c}
   \tilde{\bold G}_{A} \,&\, \tilde{\bold G}_{B} \\
   \midrule
   \tilde{\bold G}_{B} \,&\, \tilde{\bold G}_{A} \\
\end{array}\right)\!,
\end{align}
where:
\begin{align}
\tilde{{\bold G}}_{A} =\begin{pmatrix}
(N_c^2-2) & -N_c & 0 \\
-N_c & (N_c^2-2)  & 0 \\
0 & 0 & 0 
\end{pmatrix}\!,\,
\tilde{{\bold G}}_{B} =\begin{pmatrix}
-N_c & 2 & 0 \\
2 & -N_c  & 0 \\
0 & 0 & 0
\end{pmatrix}\!.
\end{align}
Due to the structure of $\tilde{\bold G}_6$, two of the matrix dimensions in this product (corresponding to those elements of ${\cal B}$ and ${\cal C}$ that do not carry an adjoint Wilson line) are superfluous and can be removed. This allows us to make the following redefinitions:
\begin{align}\label{eq:newresult3_alt}
{\cal B} \rightarrow&\,  U^{ab}_3 ({\bold B}^{ba})\\
{\cal C} \rightarrow&\, U^{cd}_3 {\bold C}^{de}\, U^{\dagger ec}_{\bar{3}},
\end{align}
where ${\bold B}^{ba}$ and ${\bold C}^{de}$ are the vectors defined in \eqref{WLvecs}. Finally, by reverting the gluon propagator decomposition performed in \eqref{factorize}, we arrive at \eqref{finalrad}.

\section{Transition terms}
\label{app:limit}
In this appendix, we recast the first three terms of \eqref{largenc1b} into a more familiar (albeit less compact) form. To do this, we recall that \eqref{largenc1b} is obtained from \eqref{largenc1} by decomposing Wilson line quadrupoles into `factorizable' and `non-factorizable' pieces (see \eqref{quad1}). However, in the large-$N_c$ limit, one may instead employ the following more widely used decomposition \cite{Blaizot:2012fh}:
\begin{align}
\left\langle Q_{12\bar{2}\bar{1}} \right\rangle_{(x_g,\bar{x}_A)}\! = &\,\left\langle S_{1\bar{1}}\right\rangle_{(x_g,\bar{x}_A)}\!\left\langle S_{\bar{2}2}\right\rangle_{(x_g,\bar{x}_A)}\nonumber\\
&+ \!\int^{x_g}_{\bar{x}_A}\!\!\diff s \left\langle S_{1\bar{1}}\right\rangle_{(x_g,s)}\!\left\langle S_{\bar{2}2}\right\rangle_{(x_g,s)}\! T(s) \left\langle S_{12}\right\rangle_{(s,\bar{x}_A)}\!\left\langle S_{\bar{2}\bar{1}}\right\rangle_{(s,\bar{x}_A)}.
\end{align}
The first term describes a quark and an antiquark propagating in a factorized double-dipole configuration, and thus rotating their color independently. Conversely, the second term corresponds to the transition from a correlated configuration to an uncorrelated one. This is realized through a single gluon exchange at time $s$, encoded in the transition amplitude $T(s)$.
The relation between the two decomposition formulas is obtained by taking the medium average of \eqref{quad1} and applying the large-$N_c$ limit:
\begin{align}
\frac{2}{N_c}\left\langle \text{Tr}\{ V_1t^aV^{\dagger}_{\bar{1}}\}\text{Tr}\{ V_{\bar{2}}t^aV^{\dagger}_{2}\}\right\rangle_{\!(x_g,\bar{x}_A)} \!\underset{N_c\uparrow}{\longrightarrow} \int^{x_g}_{\bar{x}_A}\!\!\diff s \left\langle S_{1\bar{1}}\right\rangle_{(x_g,s)}\!\left\langle S_{\bar{2}2}\right\rangle_{(x_g,s)}\! T(s) \left\langle S_{12}\right\rangle_{(s,\bar{x}_A)}\!\left\langle S_{\bar{2}\bar{1}}\right\rangle_{(s,\bar{x}_A)}\!.
\end{align}
We now apply this equivalence to the leading-$N_c$ contributions to our main result. Doing so, the first three terms of \eqref{largenc1b} become:
\begin{align}\label{eq:trans1}
\langle S_{12}S_{\bar{2}\bar{1}}\rangle_{\scriptscriptstyle (x_g,\bar{x}_A)}\!\Bigg[\!\left\langle S_{13}S_{32}S_{\bar{2}\bar{1}}\right\rangle_{\scriptscriptstyle (\bar{x}_g,x_g)}\!\!\int^{L}_{\bar{x}_g} \!\diff s \left\langle S_{\bar{2}2}\right\rangle_{\scriptscriptstyle (L,s)}\left\langle S_{3\bar{3}}\right\rangle_{\scriptscriptstyle (L,s)}\! T_1(s) \left\langle S_{32}\right\rangle_{\scriptscriptstyle (s,\bar{x}_g)} \left\langle S_{\bar{2}\bar{3}}\right\rangle_{\scriptscriptstyle (s,\bar{x}_g)}\!\left\langle Q_{\bar{3}\bar{1}13}\right\rangle_{\scriptscriptstyle (L,\bar{x}_g)}\nonumber\\
+\int^{\bar{x}_g}_{x_g} \!\diff s \left\langle S_{3\bar{1}}\right\rangle_{\scriptscriptstyle (\bar{x}_g,s)}\left\langle S_{\bar{2}2}\right\rangle_{\scriptscriptstyle (\bar{x}_g,s)} T_2(s) \left\langle S_{32}\right\rangle_{\scriptscriptstyle (s,x_g)}\!\left\langle S_{\bar{2}\bar{1}}\right\rangle_{\scriptscriptstyle (s,x_g)}\!\left\langle S_{13}\right\rangle_{\scriptscriptstyle (\bar{x}_g,x_g)}\left\langle S_{\bar{2}2}S_{3\bar{3}}Q_{\bar{3}\bar{1}13}\right\rangle_{\scriptscriptstyle (L,\bar{x}_g)}\Bigg]\nonumber\\
+\int^{x_g}_{\bar{x}_A} \!\diff s \left\langle S_{1\bar{1}}\right\rangle_{\scriptscriptstyle (x_g,s)}\left\langle S_{\bar{2}2}\right\rangle_{\scriptscriptstyle (x_g,s)} T_3(s) \left\langle S_{12}\right\rangle_{\scriptscriptstyle (s,\bar{x}_A)}\left\langle S_{\bar{2}\bar{1}}\right\rangle_{\scriptscriptstyle (s,\bar{x}_A)} \left\langle S_{13}S_{3\bar{1}}\,S_{\bar{2}2}\right\rangle_{\scriptscriptstyle (\bar{x}_g,x_g)}\!\!\left\langle S_{\bar{2}2}S_{3\bar{3}}\,Q_{\bar{3}\,\bar{1}\,1\,3}\right\rangle_{\scriptscriptstyle (L,\bar{x}_g)},
\end{align}
where we omitted the overall $\left\langle S_{12}\right\rangle_{(\bar{x}_A,x_A)}$ factor for simplicity. Also, despite having taken the large-$N_c$ limit, we choose not to explicitly factorize all correlators, in order to avoid cluttering the expression further.

In this form, \eqref{eq:trans1} is useful to examine the limit in which the antiquark propagates far from the other particles in the transverse plane (briefly discussed in \secref{sec:largenc}).
The key observation lies in the fact that \eqref{eq:trans1} contains dipoles which yield an exponential suppression in said limit, as they depend on the transverse distance between antiquark and quark both in amplitude and conjugate amplitude. This is the case, for example, of the third term, which contains a factor $\langle S_{12}\rangle_{(s,\bar{x}_A)}\langle S_{\bar{2}\bar{1}}\rangle_{(s,\bar{x}_A)} $. The other two terms are further suppressed by correlators that depend on the transverse distance between antiquark and gluon, i.e.\ $\langle S_{32}\rangle$ and $\langle S_{\bar{2}\bar{3}}\rangle$.

Conversely, this limit has virtually no effect on those contributions where the antiquark is decorrelated from the rest of the antenna. This is the case of the last term of \eqref{largenc1b}, which only gets suppressed by an overall factor shared by the rest of the contributions (i.e.\ the survival probability of the antenna, $\langle S_{12}\rangle_{(\bar{x}_A,x_A)}$), and is therefore the dominant term. As discussed in \secref{sec:limit2}, by combining this limit with the one where the quark (or, in general, the emitter prong of the antenna) is hard, one essentially recovers the BDMPS-Z spectrum times a couple of correction factors.

\bibliography{Refs}{}
\bibliographystyle{JHEP-2modlong.bst}

\end{document}